\documentclass[12pt,a4paper]{article}
\pdfoutput=1
 \usepackage{jheppub}
 \usepackage{amsmath}
 \usepackage{amsfonts}
 \usepackage{mathtools}
 \usepackage{amssymb}
 \usepackage{caption}
 \usepackage{subcaption}
 \usepackage{slashed}

 \title{Global  gravitational anomalies  and  transport}
 \author{Subham Dutta Chowdhury, Justin R. David}
\affiliation{Centre for High Energy Physics, Indian Institute of Science,\\
C. V. Raman Avenue, Bangalore 560012, India.}
\emailAdd{subham, justin@cts.iisc.ernet.in}
 
 \abstract{
 We investigate the constraints imposed by global  gravitational anomalies on parity odd 
 induced transport coefficients in even dimensions
 for theories
 with chiral fermions, gravitinos and self dual tensors. 
 The $\eta$-invariant for the large diffeomorphism corresponding 
 to the $T$ transformation on a torus   constraints 
 the coefficients in the thermal effective action  up to mod 2.  
 We show that the result obtained  for the parity odd transport for gravitinos
 using global anomaly matching  is consistent with the direct perturbative calculation.  
  In  $d=6$ we see that  the second   Pontryagin class
 in the anomaly polynomial does not contribute to the $\eta$-invariant
 which provides a topological explanation of this observation in the `replacement rule'. 
 We  then perform a direct perturbative calculation  for   the contribution of the
 self dual tensor in $d=6$ to 
 the parity odd transport coefficient  using the Feynman rules
 proposed by Gaum\'{e} and Witten.  The result for the transport coefficient 
 agrees with that obtained using matching of global anomalies. 
 }
 
\begin{document}
\maketitle
\flushbottom

\section{Introduction}

Several recent works have investigated 
the relationship between parity odd transport coefficients and 
anomalies in even dimensions 
\cite{Bhattacharyya:2007vs,Erdmenger:2008rm,Banerjee:2008th,Son:2009tf,Neiman:2010zi,Banerjee:2012iz,Banerjee:2012cr,Jensen:2012jy,Jensen:2012jh,Jain:2012rh,Valle:2012em}. 
Among these relations, the ones relating  the 
mixed gravitational anomalies or the pure gravitational anomalies to 
the appropriate parity odd transport coefficients 
are  the harder to establish. 
This is because these anomalies influence 
transport coefficients which occur at lower order in 
the derivative expansion when compared to the order they occur 
in the anomalous conservation law. 
There are three methods used to establish the 
relationship between the gravitational anomalies and the corresponding 
transport coefficients
\begin{enumerate}
 \item Direct perturbative evaluation of the Kubo formula of the transport coefficients
 using finite temperature field theory methods 
 \cite{Landsteiner:2011cp,Landsteiner:2011iq,Landsteiner:2012kd,Golkar:2012kb,Chowdhury:2015pba}.
 \item Using the method of consistency of the Euclidean vacuum 
 \cite{Jensen:2012kj,Jensen:2013rga}. 
 \item Evaluating the one loop thermal partition functions of the theory on 
 a spatial slice to obtain an  effective Chern-Simons terms which are in turn related 
 to the transport coefficients \cite{Golkar:2012kb,DiPietro:2014bca}.
\end{enumerate} 

Recently a new method has been proposed by  \cite{Golkar:2015oxw} which relies on 
matching anomalies  of   large diffeomorphisms or large gauge transformations 
to fix the Chern-Simons terms in the thermal effective action. 
One of the goals of this paper is study this approach  in more detail, 
extend the method to $d=6$ and  study   the situations in which 
there are gravitinos and self dual tensors in the theory.
A second aim of the paper is to evaluate the contribution of 
self dual tensors to parity odd transport coefficient in $d=6$ 
perturbatively using the Feynman rules for these fields put forward by 
\cite{AlvarezGaume:1983ig}. 
This calculation will also check the consistency of the result obtained using the 
matching global anomalies. 

Let us briefly summarize the method of global anomaly matching to 
determine the thermal effective action. 
Consider a $2n$ dimensional manifold
 torus $T^{2n}$.  
We will identify one of the directions of the torus to play the role 
of Euclidean time. 
Let the metric on the torus be 
$g_{\mu\nu}$ on which there exists a global diffeomorphism 
\begin{equation}
g_{\mu\nu}  \rightarrow g^{T}_{\mu\nu}.
\end{equation}
If there exists a global anomaly, the partition function of the 
theory changes by 
\begin{equation} \label{gltrans} 
Z[g_{\mu\nu}] \rightarrow 
Z[ g_{\mu\nu}^{T}]  =  e^{-i \pi \eta} Z[ g_{\mu\nu})],
\end{equation}
where $\eta$ is the  $\eta$-invariant \footnote{The details of how the topological $\eta$-invariant 
is defined and evaluated will be discussed subsequently.}
corresponding to the  global diffeomorphism. 
We will consider the  $T$-symmetry  of the torus. 
After evaluating the $\eta$-invariant, one then 
writes down a thermal effective action involving the 
components of the metric which transforms
identical to (\ref{gltrans}). 
The thermal effective action is in general 
a Chern-Simons type action considered  in all the remaining $2n-1$ directions 
excluding time. 
It is clear from (\ref{gltrans})  that coefficients in any action determined this way will 
be ambiguous up to mod 2.  
Once the effective action is obtained,   we can take 
the decompactification limit in the $2n-1$ directions and use it to 
obtain response functions  corresponding to the transport coefficients.

The contributions of chiral gravitinos to parity odd transport coefficients in the theory are 
particularly tricky to determine. 
The direct perturbative evaluation of the transport coefficient 
does not agree with that obtained using the 
method of consistency of the Euclidean vacuum \cite{Chowdhury:2015pba}.
The method of consistency of the Euclidean vacuum  
\cite{Jensen:2012kj,Jensen:2013rga} predicts that the 
transport coefficients of gravitinos is directly related to the gravitational anomaly 
by a multiplicative constant. 
For example consider the case of $d=2$,  and let 
$\lambda^{(2)} = \tilde c_{2d}T^2$ be the parity 
odd coefficient due to the presence of chiral gravitinos, then 
the method of consistency of the Euclidean vacuum predicts the 
relation 
\begin{equation}
 \tilde c_{2d} = - 8\pi^2 c_g,
\end{equation}
where $c_g$ is the gravitational anomaly due to chiral gravitinos. 
However, direct perturbative calculations  \cite{Chowdhury:2015pba}   reveal that the 
the contribution of chiral gravitinos in $d$ dimensions  to  transport is equal 
to that  of $d-1$ chiral fermions.  
In this paper we use the method  matching  global anomalies to determine the 
contribution of chiral gravitinos to transport. 
We see that result from this method is consistent  with the perturbative calculation 
 up to mod 2.  This phenomenon is consistently seen in $d=2, 6$.  

One of the observations of works of
\cite{Loganayagam:2011mu,Loganayagam:2012pz,Loganayagam:2012zg}, 
is that  second or higher Pontryagin classes in the anomaly 
polynomial of a chiral field does not contribute to 
the transport coefficient. 
This was summarized succinctly in the `replacement'  rule.  Lets recall the 
rule for the transport coefficient which is sensitive to pure gravitation anomalies in 
$d=6$.   
Consider the anomaly polynomial for pure gravitational anomalies in $d=6$, 
\begin{eqnarray}
 {\cal P}_{d=6} &=&  c_{\gamma} ( {\rm Tr} (\hat R)^2 ) ^2 + 
 c_\delta ( \frac{1}{4} {\rm Tr} ( \hat R^4)    - \frac{1}{8} {\rm Tr} 
 (\hat R^2)^2  ), \\ \nonumber
 \hat R_{ab} &=& \frac{1}{2} R_{abcd} dx^c \wedge dx^d. 
\end{eqnarray}
Note $c_\gamma$ is the coefficient  which occurs with the 
square of the first Pontryagin class while the 
$c_{\delta}$ occurs with the second Pontryagin class. 
Now let the parity odd transport coefficient determined by the
three point function of the stress tensor be parametrised as
\begin{equation}
\lambda_3^{(6)} = 9 \tilde c_g^{6d} T^4. 
\end{equation}
Then  `replacement rule' predicts the relation
\begin{equation} \label{6drr}
\tilde c_g^{6d} = - (8\pi^2)^2 c_\gamma.
\end{equation}
Note that the second Pontryagin class does not contribute to the 
transport coefficient according to this rule. 
Using the method of global anomalies to determine the 
thermal effective action  and $\lambda_3^{(6)}$, 
  we see the $\eta$-invariant corresponding to 
the $T$-symmetry of the torus for Weyl fermions, gravitinos, 
and self dual-tensors  do not receive any topological 
contribution form the second Pontryagin class for theories in $d=6$
and therefore they do not contribute to transport. 
Thus the method of global anomaly matching   provides a topological explanation for  this 
observation in the replacement rule. 
We show that the prediction (\ref{6drr}) is consistent with 
global anomaly matching for Weyl fermions, gravitinos and 
self-dual tensors. 
We  will  also explicitly verify the prediction  in (\ref{6drr})  by 
performing a direct perturbative 
evaluation of the relevant Kubo formula   using Feynman rules 
for the self dual tensor given in  \cite{AlvarezGaume:1983ig}. 
We show that indeed the contribution of the self dual tensor 
indeed agrees with that predicted by replacement rule.

The organization of the paper is as follows. 
In section \ref{rev} we briefly review the method put forward by \cite{Golkar:2015oxw} to 
set our notations. 
In  section \ref{ga2d} we study the $d=2$ case in detail for 
all the chiral fields.  Since the partition function  and  modular properties
under the $T$ symmetry of 
free chiral fields in $d=2$  are known exactly
we also compare the calculation of the $\eta$-invariant  to these results. 
We pay particular attention to the spin structure which is picked up 
by the $\eta$-invariant. 
 We extend the analysis to $d=6$ in Section \ref{ga6d}. 
 We then proceed to evaluate the transport coefficient $\lambda^{(6)}_3$ 
 for self dual tensors using the propagator  of 
 \cite{AlvarezGaume:1983ig} in 
 section \ref{gs6d}. 
 Section \ref{concl} contains our conclusions. 
 Appendix \ref{Appendix} contains the details involved in evaluating the 
 $\eta$-invariants and appendix \ref{wc6d} contains the details
 of Wick contractions manipulations and simplification of the correlators 
 in the evaluation of the the Kubo formula for 
 $\lambda^{(6)}$ for the self dual tensor in $d=6$.  Finally appendix \ref{summ}
 summarises the $\eta$ invariants corresponding to the $T^2$ transformation 
 for chiral matter in $d=2, 6, 10$.

  \section{Global anomalies and thermal effective action} \label{rev}
  
In this section we will review  the method introduced by \cite{Golkar:2015oxw} to constrain 
   thermal effective actions  using global anomalies. 
   This section will provide  the outline of the logic 
   of the method 
   using chiral fermions in $d=2$ as an example.
   This method will be implemented in detail in section \ref{ga2d}. 
   We  will also  generalize this method to theories which 
   contain  self dual and gravitinos tensors in  $d=2, d=6$

 Consider  a theory of  complex Weyl fermions  in $d=2$ on a torus $\hat T^2$. 
   Let the co-ordinates on the torus be given by 
   $(t, x)$, with the identifications
   \begin{equation}
 (t, x)\sim   (t +  2\pi n , x +  2 \pi m ).   
   \end{equation}
   Let the metric on the torus be given by  
   \begin{equation} \label{met1}
   g: \qquad\qquad   ds^2 = ( dt + a(x) dx)^2 + dx^2.
   \end{equation}
   Lets now consider the large diffeomorphism of the torus generated by the 
   transformation
   \begin{eqnarray} \label{diff1}
    \left( \begin{array}{c}
            t \\ x
           \end{array}
\right) \rightarrow 
\left(\begin{array}{cc}
       1 &  2  \\
        0 & 1 
      \end{array}
\right) \left( \begin{array}{c}
            t \\ x
           \end{array}
\right).
   \end{eqnarray}
This diffeomorphism is the $T^2$-transformation of the torus. 
From (\ref{diff1}) we see that the  transformed metric is given by 
\begin{equation}
 g^{T^2}: \qquad\qquad  ds^2 = ( dt + ( a + 2) dx)^2  + dx^2. 
\end{equation}
Thus under $T^2$ transformation of the torus we have 
$a(x) \rightarrow a(x) + 2$. 
This large diffeomorphism of the torus will be the focus of our attention. 

Consider the partition function of the theory obtained by integrating out the 
fermions defined as 
\begin{equation}
 Z[g] = \int {\cal D}{\psi } {\cal D} {\bar \psi}     \exp ( - S (  { \psi, \bar \psi, g})  ). 
\end{equation}
If the theory has a global anomaly,   the partition function picks up a phase under 
the $T^2$ transformation \cite{Witten:1985xe} which is given by 
\begin{equation}\label{partch}
 Z[g^{T^2}] = \exp (-  i \pi \eta_{1/2} ) Z[g].
\end{equation}
The $\eta_{1/2}$ invariant is defined as following. 
Consider the $3$-dimensional manifold  $\Sigma$ which  maps the metric $g$ to $g^{T^2}$ through 
a coordinate $y$. This  $3$-dimensional manifold is called the mapping torus. 
The metric is given by 
\begin{equation}\label{metsig}
 ds^2_{\Sigma}  = dy^2  + \left[ dt + ( a + 2y ) dx \right]^2 + dx^2.  
\end{equation}
This metric interpolates between the metric $g$ and $g^{T^2}$ as $y$ is dialed 
from $0$ to $1$. 
Further more from the metric it is clear that we have the identifications
\begin{equation}
 ( t, x, y ) \sim ( t - 2x, x, y+1).  
\end{equation}
Thus the torus at $y=0$ is identified with its image at $g^{T^2}$ . We will choose anti-periodic boundary 
conditions for the fermions along the time circle $t$ which will eventually be 
the thermal circle. 
Then $\eta_{1/2}$ is obtained by solving the Dirac equation in $\Sigma$ the 
3 dimensional manifold 
with this boundary condition that 
$y=0$ and $y=1$ are glued together. 
Let $\lambda$ denote the  eigen value of the Dirac operator
\begin{equation}
 {\slashed D}_{1/2}   \psi = \lambda_{1/2}  \psi.
\end{equation}
Note that $ \psi$ is a Dirac fermion in $\Sigma$. 
Then the $\eta_{1/2}$ invariant is defined by
\begin{equation}
 \eta_{1/2} = \sum_{\lambda_{1/2}}  {\rm sign} (\lambda).
\end{equation}
The subscript $1/2$ in these definitions refer to the fact that 
we are dealing with the spin-1/2 fermions. 

Evaluating the coefficient  $\eta_{1/2}$ by solving 
the Dirac equation on $\Sigma$ is not easy because of the boundary 
conditions. For the situation in which $\Sigma$ arises as 
a boundary of a manifold $B$ we can appeal to the Atiyah-Patodi-Singer  index theorem 
to obtain $\eta_{1/2}$.  Let $B$  be  a 4-dimensional manifold 
such that $\partial B = \Sigma$.
The metric on this manifold is given by 
\begin{equation}\label{metb}
 ds^2_{B} =  dr^2 + dy^2 +f(r)^2 \left[ dt + ( a + 2y ) dx \right]^2  + dx^2. 
\end{equation}
Here $r$ takes values from $0$ to $1$. $f(r)$ is a filling  function which has the property
\begin{equation} \label{limf}
 \lim_{r\rightarrow 0} f(r) = r.  
\end{equation}
Note the above limiting   behaviour of $f(r)$ together with the fact that that $t$ is periodic
with period $2\pi$ ensures that this metric does not have a conical singularity
at $r=0$.  At the boundary $r= 1$, the metric reduces to that 
of $\Sigma$. This metric is essentially that of the solid mapping torus
obtained by filling up the $t$ circle. 
The APS theorem then states that  the index of the Dirac operator on $B$ is related to 
the geometric properties of $B$ by 
\begin{eqnarray}\label{aps}
 {\rm index} ( {\slashed D}_{1/2}) (B)   = \frac{1}{24\times  8\pi^2}  \int_B {\rm Tr } ( R \wedge R )    
 +  {\cal I  }_{\partial B = \Sigma} - \frac{1}{2} {\eta}_{1/2}.
\end{eqnarray}
Here $R$ is the curvature $2$-form on $B$ and  ${\cal I }_{\Sigma}$ is 
an integral over the boundary 
of $B$  which will be 
explained in detail in the subsequent section.  At present it is sufficient 
to mention that ${\cal I }_{\Sigma}$ are corrections to the APS index theorem 
for manifolds with boundaries. 
Essentially the APS theorem offers a geometric means to 
evaluate $\eta_{1/2}$. 

Now that one has $\eta_{1/2}$ we can go back and write down an effective
action which reproduces the change in (\ref{partch}).   Let 
\begin{equation}
 Z[g] = e^{  -S_{\rm eff} }.  
\end{equation}
The metric in (\ref{met1}) has an isometry under $t \rightarrow t+\epsilon$. 
Assuming a gap in the spectrum due to the thermal boundary conditions 
in the $t$-direction we expect the 
the low lying effective action to  be independent of $t$ and will be a functional 
of only $a$. 
An effective action which reproduces the change in (\ref{partch}) is  given by 
\begin{equation}
 S{\rm eff}  =  \frac{i   \eta_{1/2} }{4}   \int a (x)  dx. 
\end{equation}
Note that under the $T^2$ transformation $a(x)  \rightarrow a(x)  +2$. 
This will ensure that  we obtain the change given in (\ref{partch}) for the partition function. 
Recall that the integral over $x$ runs from $0$ to $2\pi$. 
It is clear that this method determines the coefficient in the effective action upto 
an integer. That is if $\eta_{1/2}$ is shifted by an  even integer we would 
still be able to satisfy (\ref{partch}). 

We will implement this method systematically for chiral fermions, gravitinos and self dual tensors 
in $d=2, 6$. Our goal is to use  the effective action to  evaluate the 
parity odd transport and compare them with perturbative calculations done in 
\cite{Chowdhury:2015pba}. 
To do this we need to introduce further scalings in the metric given in (\ref{met1}) 
so that periodicities in $t$ become $\beta$ the inverse temperature. 
We must also verify that it is consistent for fermions to satisfy thermal 
boundary conditions in the geometry we are evaluating the $\eta$-invariant.  The 
periodicity in $x$ should also  be scaled to  $L$.  We will then have to take 
the $L\rightarrow \infty$ limit. 
Finally  the result  for the transport correlator should be analytically continued to Minkowski
signature. 
We will implement all these steps in detail in the subsequent sections.

\section{ Global anomalies  and transport in $d=2$}\label{ga2d}

 Before we proceed to use the APS theorem to evaluate the $\eta$ invariant for 
  the geometry $\Sigma$ given in (\ref{metsig}) we first  illustrate 
 the simple fact that the partition function of a theory defined on a  2-torus  
 which contains Weyl fermions picks up phase under the $T$ symmetry of the torus. 
 Consider free Weyl fermions on a torus with modular parameter $\tau = i \beta/L$. 
 The modular parameter is the ratio of the lengths along the time and the 
 spatial direction of the torus. 
 Let the   fermions   obey the following boundary conditions on this torus
 \begin{eqnarray}
  (A, A) :  \psi (z + 1) = - \psi(z), \qquad \psi (z+ \tau) = - \psi(z) ,  \\ \nonumber
  (P, A): \psi( z+1) = \psi(z), \qquad \psi( z+ \tau) = - \psi(z).
 \end{eqnarray}
 Both these boundary conditions result in thermal  partition functions.  For free Weyl  fermions, 
 these are easily evaluated and we obtain 
 \begin{eqnarray}
  Z_{AA}(\tau)  = \frac{\theta_3 (\tau) }{\eta(\tau) }, \qquad\qquad 
  Z_{PA}(\tau)  =  \frac{\theta_2 (\tau)}{\eta(\tau)}.
 \end{eqnarray}
 The properties of these partition function under $T$ and $S$ are given by 
 \begin{eqnarray} \label{modular}
  T^2 : Z_{AA} ( \tau + 2) = e^{ - i \frac{\pi}{6} }   Z_{AA} ( \tau + 2), \qquad  
  S: Z_{AA} ( -\frac{1}{\tau} ) = Z_{AA} ( \tau) , \\ \nonumber
  T: Z_{PA} (\tau  + 1 ) = e^{ + i \frac{\pi}{3} } Z_{PA} ( \tau), \qquad
  S: Z_{PA} ( - \frac{1}{\tau} ) = Z_{AP} (\tau) = \frac{\theta_4 (\tau)}{\eta(\tau)}.
 \end{eqnarray}
Note that it is only the  partition function $Z_{AA}$ which is 
modular invariant. This partition function has anti-periodic boundary conditions 
on both the circles. 
It returns to itself after a   $T^2 : \tau \rightarrow \tau + 2$  with a phase $e^{- i \frac{\pi}{6} }$.
We will see that the method of evaluating the phase picked up by the 
$T$ diffeomorphism of the torus using the $\eta$ invariant chooses this 
boundary conditions. 
We then will decompactify the theory along the spatial $x$ direction 
and write down an effective action which ensures that the 
partition function picks up that required phase. 

Let us now determine the change in the partition function by computing the 
$\eta$ invariant  of the $T^2$ transformation. 
Let the change in the partition function  for a theory 
containing Weyl fermions on the 2-torus under the $T^2$ transformation be given by 
\begin{equation} \label{chpact}
 Z[g^{T^2} ]  = e^{-i \pi \eta_{1/2} } Z[g] .
\end{equation}
The APS index theorem relates the $\eta$ invariant to 
the following geometric quantity on the manifold $B$. 
\begin{eqnarray}\label{indexth}
\textrm{Ind}(\slashed{D}_{\frac{1}{2}})&=& \frac{1}{24\times 8\pi^2}
\int_B \textrm{Tr}(R \wedge R) - \frac{1}{24\times 8\pi^2} 
\int_\Sigma \textrm{Tr}(\theta \wedge R)-\frac{\eta_{\frac{1}{2}}}{2}.
\end{eqnarray}
We have used
the expression for the index theorem in \cite{Eguchi:1980jx} which results in opposite 
sign for $\eta$ compared to that in \cite{Witten:1985xe}, however this is taken care 
of by introducing the negative sign for the phase shift   in (\ref{chpact}) . 
Let us recall the metric on $B$ is given by 
\begin{eqnarray}\label{metricbulk}
 ds^2_B&=& dr^2+dy^2+f(r)^2(dt+\left[a(x)+2y\right]dx)^2+dx^2.
\end{eqnarray} 
 In (\ref{indexth}),  $\theta_{\mu b}^a$ is the second fundamental form 
defined as the difference of spin connection $\omega_{\mu b}^a$ derived 
from the metric \eqref{metricbulk} and the spin connection derived from the 
product metric at the boundary which is given by 
\begin{eqnarray}\label{metricboundary}
ds^2_P&=& dr^2+dy^2+f(1)^2(dt+\left[a(x)+2y\right]dx)^2+dx^2.
\end{eqnarray}
Then
\begin{eqnarray}
\theta_{\mu b}^a &=&\omega_{\mu b}^a(B)-\omega_{\mu b}^a(B).
\end{eqnarray} 
Note that if the manifold  $B$ does not have any boundary (\ref{indexth}) 
reduces to the usual index theorem one is familiar with. 

Before we proceed to evaluate the integrals in the RHS of (\ref{indexth}) 
we will discuss the boundary conditions on the fermions. 
To begin, the boundary conditions in the $t$ direction 
are anti-periodic due to the topology of $B$. 
Note that the $(r, t)$ plane has the topology of a disc, since the $t$-circle is filled. 
Therefore, the  fermions obey anti-periodic boundary conditions along the $t$-circle. 
This is because in the $(r, t)$ plane $t\rightarrow t + 2\pi$ is just 
a rotation. 
This is identical  to the argument 
by which the fermions in $AdS_3$ obey anti-periodic boundary conditions 
along the angular directions.   Constant time slices in $AdS_3$ have the topology of a 
disc \footnote{See \cite{Maldacena:1998bw} below equation (2.1)}.  
The Dirac fermion in $\Sigma$ is  periodic  in $y$ under $y \rightarrow y+1$ as the 
metric $g$ is identified with $g^{T^2}$ under this shift. 
$y$ parametrises the direction along with the torus is mapped on to itself. 
This implies that   we must have anti-periodic boundary conditions
in $x$  because if there  are 2 or more directions in which fermions  have
periodic boundary conditions, the partition function vanishes   due to the presence of 
of fermionic zero modes. 
Thus  the partition function  is evaluated with $(A, A)$ boundary conditions in the 
$(x, t)$ directions.

We now evaluate the integrals on the LHS of (\ref{indexth}). 
Evaluating the curvature components of the metric we obtain
\begin{eqnarray}
\int_B \textrm{Tr}(R \wedge R) &=& 
- 8 \int dy dr dx dt  f'(r) \left(  f''(r)+   f(r)^3\right)  , \\ \nonumber
&=& 2(2\pi)^2 \left[   2 (f'(0)) ^2  - 2 ( f'(1) )^2   + ( f( 0) )^4 - f(1)^4 \right] .
\end{eqnarray}
In evaluating this trace we choose the orientation of the coordinates such that 
the epsilon tensor is given by $\epsilon^{txyr}=1$.  Using (\ref{limf}) we have
$f(0) = 0, f'(0) = 1$.  Substituting these values we obtain 
\begin{equation}
 \frac{1}{ 24 \times 8 \pi^2} \int_B \textrm{Tr}(R \wedge R) = 
 \frac{1}{12}  - \frac{ (f'(1))^2 }{12} - \frac{f(1)^4 }{24}. 
\end{equation}
Let us evaluate  the boundary term 
\begin{eqnarray}
 \frac{1}{24\times 8 \pi^2} \int _{\Sigma}  \textrm{Tr}(\theta \wedge R)
 = - \frac{(f'(1)) ^2 }{12}. 
\end{eqnarray}
Thus putting together the integrals on the LHS of the index theorem (\ref{indexth})
we obtain 
\begin{eqnarray}
 \frac{1}{ 24 \times 8 \pi^2}  \left ( \int_B \textrm{Tr}(R \wedge R)  
 -  \int _{\Sigma}  \textrm{Tr}(\theta \wedge R) \right) 
 =  \frac{1}{12}   - \frac{f(1)^4 }{24}. 
 \end{eqnarray}
 Substituting in (\ref{indexth}) we obtain
 \begin{equation} \label{intindex}
 \eta_{\frac{1}{2}}=\frac{1}{6} - \frac{f^4(1)}{12} + 2
 {\rm Ind}(\slashed{D}_{\frac{1}{2}}).
 \end{equation}
 However this equation indicates that  $\eta_{1/2}$ depends 
 on the filling function $f$ which clearly is not true since $\eta$ is a topological 
 invariant.  
 The reason is because the theory of Weyl fermions contains a perturbative anomaly 
 which results in the following   gravitational Chern-Simons term on $\Sigma$
 \begin{eqnarray}\label{grcsterm}
 \frac{1}{12 \times 8 \pi^2}\int_\Sigma \omega \wedge d\omega + \frac{2}{3} \omega \wedge \omega \wedge \omega &=& - \frac{ f^4(1)}{ 12}. 
\end{eqnarray}
Here $\omega$ is the spin connection in the bulk $B$ but evaluated at the boundary. 
The original applications of the $\eta$-invariant by \cite{Witten:1985xe} involved theories
which were free of perturbative anomalies,  the anomalies were 
canceled by the Green-Schwarz mechanism. Here we 
isolate the topological invariant $\eta_{1/2}$  by subtracting  the contribution of the 
gravitational Chern-Simons term \cite{Golkar:2015oxw}. 
The manifold $B$ has the topology of a solid torus, 
  the index the Dirac operator in (\ref{intindex})  is an integer. The reason is that the 
the $\eta$-invariant is defined to take into account all the terms mod 2
when the manifold $B$ has a boundary. 
Therefore   this term 
contributes to a trivial phase shift of the partition function under the 
$T$ diffeomorphism. 
Taking all this into account we obtain
\begin{equation}\label{etafermi}
\eta_{\frac{1}{2}}=\frac{1}{6}.  
\end{equation}
Thus the phase picked up  by the $T^2$ transformation is given by 
\begin{equation}
Z[g^{T^2}]  = e^{ - i  \pi \eta_{1/2}} Z[g]  = e^{ - i \frac{\pi}{6 }} Z[g].
\end{equation}
This is precisely the phase picked up the $T^2$ transformation for 
fermions with the $(A, A)$ boundary conditions   which we
evaluated by the  direct calculation in (\ref{modular}).

\subsection{Fermions}

 To obtain the thermal effective action we first need to decompactify the spatial direction. 
 Note that  so far we have worked in dimensionless units for the metric say in (\ref{met1}). 
 We first introduce dimensions by rescaling the co-ordinates  and the metric as
 \begin{eqnarray}
\tilde{x}=\frac{L x}{2\pi}, \qquad \tilde{t}=\frac{\beta t}{2\pi}, \qquad \tilde ds^2 = \frac{\beta^2}{(2\pi)^2} ds^2.
\end{eqnarray}
 Then the metric in (\ref{met1}) becomes
 \begin{eqnarray}
d \tilde{s}^2 &=& (d \tilde{t}+\tilde{a}(\tilde{x})d \tilde{x})^2+d \tilde{x}^2,
\end{eqnarray} 
where $\tilde{a}(\tilde{x})$ is defined as
\begin{eqnarray}
\tilde{a}(\tilde{x})= \frac{\beta}{L} a(x).
\end{eqnarray}
Note now the periodicities $ \tilde x $ are $\beta$ and $L$ respectively. 
Now under the $T^2$ transformation we  have
\begin{eqnarray}
T: (\tilde{t},\tilde{x}) \rightarrow (\tilde{t}+\frac{2\beta \tilde{x}}{L},\tilde{x}), 
\qquad \tilde a \rightarrow \tilde a +  \frac{2\beta}{L}.
\end{eqnarray}
We now decompactify the $\tilde x$ direction by taking $L$ to be large. 
 The modes in the $x$ direction then 
become a continuum, but we expect the effective action  to  still retain the shift by the 
phase $e^{- i \frac{\pi}{6} }$ under the $T^2$ diffeomorphisms. 
The metric in (\ref{met1}) is flat, there is no background curvature, therefore  the effective
action can only depend on $\tilde a$. An action which satisfies the required 
condition of the phase shift is given by 
\begin{eqnarray}
S_{{\rm eff}} = \frac{i \pi}{12 \beta } \int \tilde{a}(\tilde{x}) d\tilde{x}\qquad Z[g]=e^{- S_{{\rm eff}} }. 
\end{eqnarray}
Writing this partition function in  momentum space we obtain 
\begin{equation} \label{partfn}
Z = \exp ( -\frac{i\pi}{12} \hat  a (0) ). 
\end{equation}
Here $\hat  a(0)$ is the Fourier transform  at $k=0$ \footnote{The Fourier transform is 
defined by $  \tilde a(x) = \int \frac{dk }{2\pi} \hat a(k) e^{ - i k x} $. }. 
We can  now
obtain the one point function of the stress tensor $\langle T^{\tilde \tau \tilde x} \rangle$. 
Note that since $\hat  a $ is the  $\tilde g_{\tilde t \tilde x}$ component,  the 
one point function of the stress tensor  by  \footnote{This definition of 
stress tensor is in accordance with \cite{Banerjee:2012iz} for the mostly positive signature.} 
\begin{equation}
 \langle T^{\tilde t \tilde x} (p)  \rangle = 
 \frac{1}{\sqrt{g} } \frac{\delta \ln Z}{\delta g^{\tilde t \tilde x} } 
 =  \frac{ \delta \ln  Z}{ \delta \hat a (p) }. 
\end{equation}
Evaluating this for the partition function given in (\ref{partfn})  we obtain 
\begin{eqnarray}
\langle T^{\tilde{t} \tilde{x}}(p) \rangle
&=& -\frac{i\pi}{\beta 12} 2\pi \delta(p), \nonumber\\
&=& -\frac{i\pi}{\beta^2 12} 2\pi \beta \delta(p).
\end{eqnarray}
Note that due the definition of the Fourier transform functional differentiation in Fourier space 
picks up a factor $2\pi \delta(p)$. 
We can now go over to Minkowski space by analytical continuation of $ t' = -i \tilde t$. 
This results in 
\begin{eqnarray}\label{onepointfn}
\langle T^{{t}' \tilde{x}'}(p) \rangle &=& \frac{ \langle T^{\tilde{t} \tilde{x} }(p)\rangle}{i}, \nonumber\\
&=&-\frac{\pi}{\beta^2 12} 2\pi \beta \delta(p).
\end{eqnarray}
The transport coefficient 
$\lambda^{(2)}$ which occurs in the constitutive relation for the stress tensor is obtained by 
evaluating the one point function $ - \langle T^{{t}' \tilde{x}'}(p) \rangle $
and then stripping out the $2\pi \beta \delta(0)$ which 
occurs in the overall momentum conservation of the correlators in the Kubo formula, see 
\cite{Chowdhury:2015pba}
for a discussion. 
We therefore get 
\begin{eqnarray}
\lambda^{(2)} 
&=& \frac{\pi}{12 \beta^2}.  \nonumber\\
\end{eqnarray}
This coincides with the expression obtained using perturbative calculations in 
\cite{Chowdhury:2015pba} as well as the result using the 
replacement rule  \cite{Jensen:2012kj}. 

\subsection{Chiral bosons}

The chiral boson or the self dual tensor in $d=2$ is dual to the Weyl fermion by bosonization. 
Therefore we expect the same result for the transport coefficient. 
Lets verify this by evaluating the $\eta_S$ for the self dual tensor in $d=2$. 
The $\eta$ invariant for the self dual tensor can be determined using the 
APS index formula for self dual tensors. 
To be general and also relate it to the expressions in 
\cite{Witten:1985xe} we quote the result for arbitrary dimensions. 
\begin{eqnarray} \label{hirz}
\frac{\sigma (B) }{8} &=& \frac{1}{8} L(R) - {\cal I}_\Sigma(R)   + \frac{\eta_S}{2}.
\end{eqnarray}
Here $L$ is the Hirzebruch polynomial constructed out of the 
curvature tensor, $\sigma$ the Hirzebruch signature of $B$.  ${\cal I}$ is  a
boundary term which will be defined later. 
Note that our definition of $\eta_S$ is $1/4$ the definition used in 
\cite{Witten:1985xe}. 
The change in the partition function is given by $Z[g^T] = e^{ - i \pi \eta_S} Z[g]$.
Recently the expression in (\ref{hirz}) has been refined
by \cite{Monnier:2011rk,Monnier:2013kna}.  
The term $\frac{\sigma (B) }{8}$ on the LHS  of the equation in 
(\ref{hirz})  is replaced by a $\lambda\wedge \lambda$ where 
$\lambda$ is a $2k +2$ form for a self dual field in $4k +2$ dimensions \footnote{
We thank Samuel Monnier for bringing the references 
\cite{Monnier:2011rk,Monnier:2013kna} to our attention and for explaining the 
refined formula for the $\eta$ invariant to us. See section {4.3}  of \cite{Monnier:2011rk} for the index 
theorem without gauge fields and \cite{Monnier:2013kna} for the 
index theorem with gauge fields. }. Below we will 
carry our arguments for the version of the index theorem for 
self dual tensors in \cite{Witten:1985xe} as well as 
provide the arguments for the refined formula for the $\eta$ invariant
 given in \cite{Monnier:2011rk}. 
Let us now substitute the appropriate polynomials for $B$ of dimension $4$
in the APS index theorem. 
We obtain
\begin{eqnarray}
\frac{\sigma (B) }{8} &=&
- \frac{1}{24\times 8\pi^2}
\int_B \textrm{Tr}(R \wedge R) + \frac{1}{24\times 8\pi^2}
\int_\Sigma \textrm{Tr}(\theta \wedge R)+ \frac{\eta_S}{2}.\nonumber\\
\end{eqnarray}
The evaluation of the integrals proceeds exactly as in the case of the 
fermions. We obtain the relation
\begin{equation}
\eta_S  = \frac{1}{6}  - \frac{f^4(1)}{12}  + \frac{\sigma(B)}{4}.
\end{equation} 
Note that again there is a contribution to $\eta_S$ which is removed by 
subtracting the gravitational  Chern-Simons term in (\ref{grcsterm}). 
We now use the fact that for a solid torus  $B$ the 
Hirzebruch   signature is  a multiple of $8$, again this is 
because the $\eta$ invariant is defined to take  in account of all terms mod 2. 
For the refined global anomaly formula \cite{Monnier:2011rk}, it is in fact not necessary to make 
assumptions regarding the Hirzebruch signature of $B$. 
The term $\sigma(B)/8$ is replaced by $\lambda\wedge\lambda$ where
$\lambda$ is a 2-form. 
Since in $d=2$  the manifold is a spin manifold  we can take the manifold $B$ to 
have a spin structure compatible with $T^2$. In this situation one 
can show that $\lambda$ can be chosen to vanish \footnote{
We thank Samuel Monnier for this explanation.}. 
 Therefore  to mod 2 we write
\begin{equation}
\eta_S  = \frac{1}{6}. 
\end{equation}
This result is identical to that obtained in (\ref{etafermi}) for Weyl fermions. 
Therefore the result for the transport coefficient  $\lambda^{(2)}$ 
in theories with a single self dual tensor is identical  to that of
a single Weyl fermion.

\subsection{Gravitinos} 

Though there are no physical gravitinos in $d=2$, we can study 
the `gravitino like' theory. The 
 gravitino action consists  essentially  of the spin 3/2 operator acting on 
the gravitino, the ghosts in the gauge fixing 
procedure are taken into account by   subtracting the contribution of a Weyl fermion 
Transport coefficient for this theory is evaluated perturbatively 
in \cite{Chowdhury:2015pba}. The results are inconsistent  with that obtained using the 
argument involving the consistency of the Euclidean vacuum. 
We will show the method of global anomaly matching 
is consistent with the perturbative evaluation in \cite{Chowdhury:2015pba} to mod 2.

The index theorem for   gravitinos   on closed manifolds   is given by 
\cite{AlvarezGaume:1983ig,AlvarezGaume:1984dr}
\begin{equation}
{\rm Index}( \slashed{D}_{3/2} (B) ) = \int_B  \hat A( B) \left( 
{\rm Tr} e^{ i R/2\pi }  - 1 \right). 
\end{equation}
Expanding the curvature polynomials  $\hat A(B)$ for the case when the manifold
is of $4$ dimensions we obtain 
\begin{equation}
 \textrm{Ind}(\slashed{D}_{\frac{3}{2}} (B) )= -  \frac{23}{24\times 8\pi^2}
\int_B \textrm{Tr}(R \wedge R).
\end{equation}
Therefore following \cite{Witten:1985xe} \footnote{
In \cite{Witten:1985xe} the index theorem 
for gravitinos is written as the difference of curvature 
polynomial appropriate for  only the spin $3/2$ field  and the curvature 
polynomial for fermions. We have combined the two polynomials, so that
the end result is that for the `physical' gravitino.} , the index theorem for 
manifolds with boundary is given by 
\begin{equation}
\textrm{Ind}(\slashed{D}_{\frac{3}{2}} (B) )= -  \frac{23}{24\times 8\pi^2}
\int_B \textrm{Tr}(R \wedge R) + \frac{23}{24\times 8\pi^2} 
\int_\Sigma \textrm{Tr}(\theta \wedge R)-\frac{\eta_{\frac{3}{2}}}{2}.
\end{equation}
Again our definition of $\eta_{3/2}$ is of opposite sign compared to that in 
\cite{Witten:1985xe} since in our notation the change in the 
effective action is given by $Z[g^T] \rightarrow e^{- i \eta_{3/2} } Z[g]$. 
The integrals are identical to the case of the Weyl fermions  and proceeding 
From \eqref{indexth}, we see that the only difference 
between the index theorem for gravitinos is the pre factor in 
front of Tr$(R\wedge R)$ and Tr$(\theta \wedge R)$. Proceeding similarly as before we obtain 
\begin{eqnarray}
\eta_{3/2} &=&\frac{-23}{6} - \frac{23 f^4(1)}{12} + 2 {\rm Ind} (\slashed{D}_{\frac{3}{2}} (B) ).  
\end{eqnarray} 
Again the dependence on the filling function $f(1)$ can be removed by adding 
a Chern-Simons term. 
The index of the spin $3/2$ operator on $B$ is an integer
 since the manifold $B$ has the topology of the solid torus.  The $\eta$ invariant 
 takes into account of all terms mod $2$. 
Therefore we obtain that the $\eta$ invariant for gravitinos to be 
\begin{eqnarray}
 \eta_{3/2} &=&  \frac{-23}{6}  \quad  {\rm mod}\;  2 , \\ \nonumber
 &=& \frac{1}{6} \quad  {\rm mod}\;  2.
\end{eqnarray}
We see that   upto mod $2$ the $\eta$ invariant for  Weyl gravitinos is identical to that
of the Weyl fermions. 
This result is consistent with the  direct perturbative calculations 
of the transport coefficients for gravitinos done in \cite{Chowdhury:2015pba}.
The perturbative calculations also show that 
 $\lambda^{(2)}$ for gravitinos is  identical to that 
of Weyl fermions in $d=2$. 
Therefore we conclude that in $d=2$, the transport coefficient for 
gravitinos obtained from  matching  global anomalies 
is consistent with perturbative calculations to mod 2.

\section{Global anomalies  and transport in  $d=6$}\label{ga6d}

In $d=6$ we start with the following  metric $g$  on $\hat T^6$. 
\begin{eqnarray} \label{6dgmet}
  ds^2 = ( dt + a_1( a) da + a_2 (b) dz + a_3 ( y) dx)^2 +  dx^2+dz^2+da^2+db^2+dy^2. 
\end{eqnarray}
Note that now the co-ordinate $a$ will play the role of $x$ in the previous section, this convention
has been chosen so that it agrees with that in \cite{Chowdhury:2015pba}.  
The co-ordinates satisfy the periodicity
\begin{eqnarray}
 t \sim t + 2\pi,  \quad  a \sim a + 2\pi,  \quad b \sim b +2\pi,  \\ \nonumber
 x \sim x+ 2\pi, \quad y \sim y + 2\pi, \quad  z \sim z + 2\pi.
\end{eqnarray}
We choose  anti-periodic boundary conditions for the  fermions in 
all the directions along the torus. 
Consider the $( x, y) $ plane: since we have $(A, A)$ boundary conditions on this plane, 
an  allowed non-trivial field configuration for the metric component $a_3$ is given by 
\begin{equation}\label{q1}
 a_3(y)  =   2n  \frac{y}{2\pi} , \qquad  n \in \mathbb{Z}.
\end{equation}
This ensures that the $a_3 \rightarrow a_3 + 2n$ under $ y \rightarrow y + 2\pi$.
Thus we have a $ T^2$ transformation in the torus along the $(x, y)$ plane. 
Therefore  the boundary conditions $(A, A)$ remain invariant  in these directions. 
Similarly in the $(z, y)$ plane,
 to preserve the $(A, A)$ boundary conditions in these directions  we 
consider the non-trivial field configuration 
\begin{equation}\label{q2}
 a_2(b ) = 2m  \frac{b}{2\pi}, \qquad m \in \mathbb{Z}.
\end{equation}
Thus the fields $a_3(y)$ and $a_2(b)$ have non-trivial windings along the compact direction. 
The metric component $a_1(a)$ will play the role of  
field $a(x)$  for the $d=4$ case discussed in the 
previous sections. 
We wish to consider the metric related to the one in \ref{6dgmet} by the $T^2$  
diffeomorphism  given by 
\begin{eqnarray} \label{6dgmett}
   ds^2 &=& ( dt + ( a_1( a)+ 2)  da + a_2 (b) dz + a_3 ( y) dx)^2  \\ \nonumber
          & & \qquad\qquad +  dx^2+dz^2+da^2+db^2+dy^2.
\end{eqnarray}
The  7-dimensional mapping torus $\Sigma$ which 
interpolates between (\ref{6dgmet}) and (\ref{6dgmett}) 
is given by 
\begin{eqnarray}
 ds^2_{\Sigma}
 &=& du^2+(dt+\left[a_1(a)+ 2 u\right]da+a_2(b)dz+a_3(y)dx)^2 \\ \nonumber
  & & \qquad\qquad +dx^2+dz^2+da^2+db^2+dy^2. 
\end{eqnarray}
Here the coordinate $u$ interpolates between the torus (\ref{6dgmet}) and 
the one related to it by $T^2$ diffeomorphism in  (\ref{6dgmett})  as $u$ runs
from $0$ to $1$. 
Therefore we have the identifications
\begin{equation}
 ( t, a, u, b , z, x, y ) \sim ( t- 2a, a, u+1, b, z, x, y ). 
\end{equation}
 Filling up the time circle we get,
\begin{eqnarray}\label{metric6d}
 ds^2_B &=& dr^2+ du^2+ f(r)^2(dt+\left[a_1(a)+ 2 u\right]da+a_2(b)dz+a_3(y)dx)^2  \\ \nonumber 
 & &  \qquad \qquad  +dx^2+dz^2+da^2+db^2+dy^2. 
\end{eqnarray}
The filling function satisfies the condition
\begin{equation}
\lim_{r\rightarrow 0} f(r)  = r ,
\end{equation}
for the absence of conical singularities. 
The product metric at the boundary  $r=1$ is defined to be,
\begin{eqnarray}\label{metric6dp}
 ds^2_P &=& dr^2+ du^2+ f(1)^2(dt+\left[a_1(a)+ 2 u\right]da+a_2(b)dz+a_3(y)dx)^2 \\ \nonumber
 & &  \qquad\qquad +dx^2+dz^2+da^2+db^2+dy^2.
\end{eqnarray}
The metric  on $B$  given in  (\ref{metric6d}) can now be used  to 
evaluate the $\eta$ invariant for the $T^2$ diffeomorphism using the APS index theorem. 

The index theorem  for  the Dirac operator on $B$ is given by 
\begin{eqnarray}\label{indexth6d}
\textrm{Ind}(\slashed{D}_{\frac{1}{2}} (B))&=&
 \frac{-1}{6!}\int_B \left(\frac{p_2({{R}})}{2}-\frac{7}{8}p_1({{R}})^2\right)-\frac{1}{6!\times (2\pi)^4} \int_\Sigma \left(\frac{1}{8}\textrm{Tr}(\theta \wedge R\wedge R\wedge R)\right.\nonumber\\
&&\left. -\frac{1}{16}\textrm{Tr}(\theta \wedge R)\textrm{Tr}(R\wedge R)+\frac{7}{32}\textrm{Tr}(\theta \wedge R)\textrm{Tr}(R\wedge R)\right)-\frac{\eta_{1/2} }{2}. 
\end{eqnarray}
where  the Pontryagin classes are defined by 
\begin{eqnarray}
p_2({{R}})&=&\frac{-1}{(2\pi)^4}\left(\frac{1}{4}\textrm{Tr}(R \wedge R\wedge R\wedge R)-\frac{1}{8}\textrm{Tr}(R \wedge R)\textrm{Tr}( R\wedge R)\right),\nonumber\\
p_1({{R}})&=&\frac{-1}{(2\pi)^2} \frac{\textrm{Tr}(R \wedge R)}{2},
\end{eqnarray}
and $\theta$ is the difference of the spin connections of $B$ and the product metric $P$. 
\begin{equation}
\theta_{\mu b} ^{a} = \omega_{\mu b} ^a(B) - \omega_{\mu b} ^a(P). 
\end{equation}
Essentially the index theorem in  (\ref{indexth6d})   is  
for  a closed $8$-manifold with the corrections due to the presence of a boundary.
These corrections are obtained by replacing a single  $R$ in the anomaly polynomial 
by $\theta$ \cite{Witten:1985xe}. Once  $\eta_{1/2}$ is obtained  the change in the 
partition function is given by 
\begin{equation}
Z[g^{T^2}] = e^{- i \eta_{1/2}} Z[g].
\end{equation}
We proceed to evaluate the integrals occurring on the RHS of (\ref{indexth6d}).

\subsubsection*{Contribution from the second Pontryagin class}

We first  show that all the contributions from the second Pontryagin 
class to the index theorem in (\ref{indexth6d}) is identical  to a Chern-Simons term
on the boundary $\Sigma$. 
Consider the term involving the second  Pontryagin  class in (\ref{indexth6d}) 
along with the associated boundary term which is given by 
\begin{eqnarray}
I_{p_2}&=& \frac{-1}{6!}\int_B \left(\frac{p_2({\cal{R}})}{2}\right)+\frac{-1}{6!\times (2\pi)^4}
\int_\Sigma \left(\frac{1}{8}\textrm{Tr}(\theta \wedge R\wedge R\wedge R)\right.\nonumber\\
&&\left. -\frac{1}{16}\textrm{Tr}(\theta \wedge R)\textrm{Tr}(R\wedge R)\right).\nonumber 
\end{eqnarray}
Substituting the definition of the second Pontryagin class we obtain 
\begin{eqnarray}
 I_{p_2} &=& \frac{1}{6!(2\pi)^4}\int_B\left(\frac{1}{8}\textrm{Tr}(R \wedge R\wedge R\wedge R)
-\frac{1}{16}\textrm{Tr}(R \wedge R)\textrm{Tr}( R\wedge R)\right)\nonumber\\
&&+\frac{-1}{6!\times (2\pi)^4} \int_\Sigma \left(\frac{1}{8}\textrm{Tr}(\theta \wedge R\wedge R\wedge R) 
-\frac{1}{16}\textrm{Tr}(\theta \wedge R)\textrm{Tr}(R\wedge R)\right).
\end{eqnarray}
Each of these  integrals have been evaluated in (\ref{curvint}).
On substituting the results for the integrals we obtain 
\begin{eqnarray} \label{ip2}
I_{p_2} 
&=& \frac{-1}{4\times 6!} \int db dy  a_2'(b) a_3'(y) \left\{3 f(1)^8 
\left[a_2'(b)^2+ a_3'(y)^2+a_2'(b)^2 \left(1+a_3'(y)^2\right)\right]\right.\nonumber\\
&&\left. + 4f(1)^4 f'(1)^2 \left(1+a_2'(b)^2+a_3'(y)^2\right)\right\}.
\end{eqnarray}
Note that in the integral $I_{p_2}$, all  terms  depend on the filling function $f$, 
therefore we expect the entire contribution not to contribute to 
the topological $\eta$ invariant. 
We will now show that the entire  contribution  of $I_{p_2}$ can be be accounted by a
Chern-Simons term on the boundary $\Sigma$.  
Let us define the Chern-Simons form  on $\Sigma$
\begin{eqnarray}\label{cs1int}
 {\cal I}_{CS_1} &=& \frac{1}{(2\pi)^4}  \frac{1}{8}
\left[ \textrm{Tr} (\omega \wedge d\omega\wedge d\omega\wedge d\omega) +
\frac{8}{5}\textrm{Tr}( d\omega \wedge d\omega \wedge \omega \wedge \omega \wedge \omega )\right.  \\ 
& & 
+\frac{4}{5}\textrm{Tr} (d\omega \wedge \omega \wedge d\omega  \wedge \omega \wedge \omega)+ \nonumber\\
&& + \frac{4}{7}
\textrm{Tr}( \omega \wedge \omega \wedge \omega \wedge \omega \wedge \omega \wedge \omega
\wedge \omega )
+ 2\textrm{Tr}( d\omega \wedge \omega\wedge \omega\wedge \omega\wedge \omega\wedge \omega)
\nonumber\\
& &  \left. -\frac{1}{16} \textrm{Tr}(\omega \wedge d\omega 
+\frac{2}{3} \omega \wedge \omega \wedge\omega )\textrm{Tr}(R\wedge R) \right] .
\end{eqnarray}
It is can be verified that 
\begin{equation}
 -\frac{1}{6!} p_2(R) = \frac{1}{6!} d \left( {\cal I}_{CS_1} \right). 
\end{equation}
We now integrate the Chern-Simons form ${\cal I}$ on the boundary $\Sigma$. 
The integrals of each of the terms occurring in ${\cal I}$ is given in (\ref{csintegrals}). 
Putting all the terms together  we obtain
\begin{eqnarray}\label{ics1}
 I_{CS_1} &=& \frac{1}{6!} \int_\Sigma {\cal I}_{CS_1}, \nonumber\\ \nonumber
 &=& \frac{-1}{4\times 6!} \int db dy  a_2'(b) a_3'(y) \left\{3 f(1)^8 
\left[a_2'(b)^2+ a_3'(y)^2+a_2'(b)^2 \left(1+a_3'(y)^2\right)\right]\right.\nonumber\\
&&\left. + 4f(1)^4 f'(1)^2 \left(1+a_2'(b)^2+a_3'(y)^2\right)\right\}.
\end{eqnarray}
 In performing this integrals note that the spin connection $\omega, d\omega$ is evaluated at the boundary $r=1$, 
 The curvature forms in the last term of (\ref{cs1int}) is the curvature of the bulk  metric 
 evaluated at the boundary. 
 
 It is indeed remarkable that the contribution of the second Pontryagin class to the 
 APS index theorem given in (\ref{ip2}) coincides precisely with the value of the Chern-Simons form 
 in (\ref{ics1}). Therefore  we can completely remove the contribution of 
 the second Pontryagin class by subtracting out  the Chern-Simons term  (\ref{ics1}) at the boundary. 
As in the case of $d=2$ studied in the earlier section, this operation ensures that 
we pick up only the purely topological terms in the $\eta$ invariant. 
Thus we conclude the the second Pontryagin class does not contribute to the 
purely topological terms in the $\eta$ invariant and therefore will not contribute to 
transport.  This was observed in the `replacement rule' 
\cite{Loganayagam:2012zg,Jensen:2012kj,Jensen:2013rga} as well as the holographic calculations
of \cite{Azeyanagi:2013xea}.

\subsubsection*{Contributions from square of the first Pontryagin class}

On the RHS of the index theorem  (\ref{indexth6d}) 
the contributions due to $p_1^2$ is given by 
\begin{eqnarray}
I_{p_1}&=&\frac{7}{8\times 6!}\int_X \left(p_1({{R}})^2\right)+\frac{-7}{32\times 6!\times (2\pi)^4} \int_\Sigma \left(\textrm{Tr}(\theta \wedge R )\textrm{Tr}( R\wedge R)\right),\nonumber\\
&=&\frac{7}{32\times 6!\times(2\pi)^4}\int_X \left(\textrm{Tr}(R\wedge R )\textrm{Tr}( R\wedge R)\right)-\frac{7}{32\times 6!\times (2\pi)^4} \int_\Sigma \left(\textrm{Tr}(\theta \wedge R )\textrm{Tr}( R\wedge R)\right).\nonumber\\
 \end{eqnarray}
 Substituting from (\ref{curvint}) for the curvature integrals we obtain 
 \begin{eqnarray}\label{intp11}
 I_{p_1} 
 &=&\frac{-7}{32\times 6!} f(1)^4 a_2'(b) a_3'(y) \left\{-16 f'(1)^2 \left(a_2'(b)^2+a_3'(y)^2+1\right) \right. \\
&&\left. -f(1)^4 \left[11 a_2'(b)^2 \left(a_3'(y)^2+1\right)+5 a_2'(b)^4+5 a_3'(y)^4+11 a_3'(y)^2+5\right] +48\right\}. \nonumber
  \end{eqnarray}
 Note that this integral contains terms which depend on the filling function $f$ as well 
 as the pure topological term which arises from the last term in (\ref{intp11}). 
 We will again show that all terms that depend on the filling function 
 can be canceled by a Chern-Simons term evaluate at the boundary $\Sigma$. 
 From the appendix  we have the identity
 \begin{equation}
 p_1^2 (R) =  \frac{1}{4 (2\pi)^4 }
 d ( \textrm{Tr}(\omega \wedge d\omega +\frac{2}{3} \omega \wedge \omega \wedge\omega )\textrm{Tr}(R\wedge R).
 \end{equation}
 Therefore we consider the Chern-Simons term
 \begin{eqnarray}
I_{CS_2}&=&\frac{7}{32\times 6!\times(2\pi)^4} \int _\Sigma \textrm{Tr}(\omega \wedge d\omega +\frac{2}{3} \omega \wedge \omega \wedge\omega\textrm{Tr}(R\wedge R).
 \end{eqnarray}
 Substituting for the spin connection at the boundary and the bulk curvature, but evaluated
 at the boundary we obtain 
 \begin{eqnarray} \label{ics2}
 I_{CS_2} 
 &=&\frac{-7}{32\times 6!} f(1)^4 a_2'(b) a_3'(y) \left\{-16 f'(1)^2 \left(a_2'(b)^2+a_3'(y)^2+1\right) \right. \\
&&\left. -f(1)^4 \left[11 a_2'(b)^2 \left(a_3'(y)^2+1\right)+5 a_2'(b)^4+5 a_3'(y)^4+11 a_3'(y)^2+5\right] \right\} .\nonumber
\end{eqnarray}
Note the absence of the last term of (\ref{intp11}) in (\ref{ics2}).

Now using the results in (\ref{ip2}) , (\ref{ics1}), (\ref{intp11}) and (\ref{ics2}) we can write the 
index theorem  in (\ref{indexth6d}) as
\begin{eqnarray}
\frac{\eta_{1/2}}{2} &=& - \frac{7 }{ 480 (2\pi)^2 }
\int db dy da dz a_2'(b)a_3'(y)+I_{CS1}+I_{CS2} -  {\rm Ind} ( \slashed D_{1/2} (B)).
\end{eqnarray}
Again note that removing the Chern-Simons terms we obtain a purely topological 
$\eta$ invariant, also  the index of the Dirac operator is an integer. 
Therefore 
we obtain that the shift of the 
phase in the path integral under the $T^2$ transformation, $a_1(a) \rightarrow a_1(a) + 2$
 is given by 
\begin{eqnarray}
\eta_{1/2} =   - \frac{7 }{ 240 (2\pi)^2 }
\int db dy dx dz a_2'(b)a_3'(y)    \qquad  {\rm mod }\;  2.
\end{eqnarray}
Due to the quantization  conditions (\ref{q1}) and (\ref{q2}) we obtain 
\begin{equation}
\eta_{1/2} = - \frac{ 7 nm }{60} \qquad  {\rm mod }\;  2.
 \end{equation}
 An effective action which reproduces this phase shift is given by 
 \begin{equation} \label{6deff}
 S_{{\rm eff}} =  - \frac{i 7 \pi  }{ 480 ( 2\pi)^3} \int da db dx dy dz a_1(a)  a_2'(b)a_3'(y). 
 \end{equation}
 Note that under $a_1 \rightarrow a_1+2$ the phase shift from this effective action is 
 given by $e^{ -i \pi \eta_{1/2}}$. 
 This effective action can formally be written as a Chern-Simons form 
 by introducing the graviphoton field $A = A_\mu dx^\mu$ as 
 \begin{equation}
 S_{{\rm eff}} = - \frac{i  7 \pi  }{ 960 ( 2\pi )^3} \int A \wedge d A \wedge d A.
 \end{equation}
 
 \subsection{Fermions}
 
 To take the decompactification limit and  to introduce the temperature we 
 resale the coordinates as
 \begin{eqnarray}
\tilde{a}=\frac{L_a a}{2\pi}, \qquad \tilde{t}=\frac{\beta t}{2\pi}, \qquad \tilde{z}=\frac{L_z a}{2\pi}, \nonumber\\
\tilde{x}=\frac{L_x x}{2\pi}, \qquad \tilde{y}=\frac{L_y a}{2\pi}, \qquad \tilde{b}=\frac{L_b b}{2\pi}.  
\end{eqnarray}
After introducing dimensions by  rescaling the metric using
\begin{equation}
\tilde ds^2 = (\frac{\beta}{2\pi} ) ^2 ,
 \end{equation}
 the metric in (\ref{6dgmet}) becomes
 \begin{eqnarray}
d\tilde{s}^2&=& (d\tilde{t}+\tilde{a}_1(\tilde{a})d \tilde{a}+\tilde{a}_2(\tilde{b})d\tilde{z}+\tilde{a}_3(\tilde{y})d\tilde{x})^2 \\ \nonumber
& & \qquad\qquad+ (\frac{\beta}{L_x})^2 d\tilde{x}^2+(\frac{\beta}{L_z})^2 
d\tilde{z}^2+(\frac{\beta}{L_a})^2 d\tilde{a}^2+(\frac{\beta}{L_b})^2d\tilde{b}^2+
(\frac{\beta}{L_y})^2d\tilde{y}^2, 
\end{eqnarray}
where $\tilde{a_1}, \tilde{a_2}, \tilde{a_3}$ are defined as,
\begin{eqnarray}
\tilde{a_1}= \frac{\beta}{L_a} a_1, \qquad \tilde{a_2}= \frac{\beta}{L_z} a_2, \qquad\tilde{a_3}= \frac{\beta}{L_x} a_3.
\end{eqnarray}
After these change of variables, the effective action in (\ref{6deff}) becomes 
\begin{equation}
S_{{\rm eff}} = 
\frac{-i7\pi}{\beta^3 480} \int d\tilde{b} d\tilde{y}  d\tilde{a} d\tilde{z} d\tilde{x} \tilde{a}_1(\tilde{a}) \tilde{a}_2'(\tilde{b}) \tilde{a}_3'(\tilde{y}).
\end{equation}
We decompactify the spatial directions and then  write the action in Fourier space, 
we obtain
\begin{equation}
S_{{\rm eff}} = 
\frac{-i7\pi}{\beta^3 480} \int \frac{d^5 p d^5 k}{((2\pi)^5)^2}(ik^b ip^y)\tilde{a}_1(-p-k) \tilde{a}_2(k) \tilde{a}_3(p).
\end{equation}
The transport coefficient $\lambda_3^{(6)}$ which is sensitive to the pure gravitational 
anomaly is defined by the following Kubo formula \cite{Chowdhury:2015pba}
\begin{eqnarray}
\lambda^{6}_3 &=& -\frac{3 \langle T^{ta}(-p-k)T^{tx}(p)T^{tz}(k) \rangle}{2( ip^y) (ik^b) }, \nonumber\\
&=& -\frac{3i\langle T^{\tau a}(-p-k)T^{\tau x}(p)T^{\tau z}(k) \rangle}{2 ( ip^y)(ik^b) },\nonumber\\
&=& -\frac{3i}{(2ip^y)(ik^b)} \frac{\delta^3 \ln Z}{\delta g_{\tau a} \delta g_{\tau x}  \delta g_{\tau z} }.
\end{eqnarray}
In the second line of the above equation we have analytically continued to 
Euclidean correlators  using $t = - i \tau$. In the last line we have written the correlator 
in terms of derivatives on the partition function. 
Using $\ln Z = - S_{{\rm eff}}$  and identifying 
$\delta g_{\tau a} = \delta a_1,  \delta g_{\tau z} = \delta a_2, \delta g_{\tau x} = \delta a_3$
 we obtain
\begin{eqnarray}
\lambda^{6}_3
&=&  -\frac{3i}{2(ip^y)(ik^b)} \frac{(i7\pi) 
( ip^y)(ik^b)}{\beta^4 480} (2\pi)^5 \beta \delta(0), \nonumber\\
&=& \frac{7\pi }{320 \beta^4}  \times (2\pi)^5 \beta \delta(0).
\end{eqnarray}
In the last line we have factored out the terms which are due to the overall momentum 
conservation.  
Therefore we obtain
\begin{equation}
\lambda^{6}_{3(1/2)}  = \frac{7\pi }{320 \beta^4}. 
\end{equation}
This result coincides with the one obtained in \cite{Chowdhury:2015pba} using perturbation theory
at one loop \footnote{Note the first term in equation (4.39) of  \cite{Chowdhury:2015pba}. }. 

\subsection{Gravitinos}

We will now show that the constraints obtained for the thermal effective action 
for gravitinos 
using global anomalies is consistent with the result for the 
transport coefficient $\lambda_3^{(6)}$ obtained using perturbation theory in 
\cite{Chowdhury:2015pba}. 
The APS index theorem for gravitinos is given by 
\begin{eqnarray}
\textrm{Ind}(\slashed{D}_{\frac{3}{2}})&=&\frac{-1}{6!}\int_B \left(\frac{245 p_2({\cal{R}})}{2}-\frac{275}{8}p_1({\cal{R}})^2\right)-\frac{1}{6!\times (2\pi)^4} \int_\Sigma \left(\frac{245}{8}\textrm{Tr}(\theta \wedge R\wedge R\wedge R)\right.\nonumber\\
&&\left. -\frac{245}{16}\textrm{Tr}(\theta \wedge R)\textrm{Tr}(R\wedge R)+\frac{275}{32}\textrm{Tr}(\theta \wedge R)\textrm{Tr}(R\wedge R)\right)-\frac{\eta_{\frac{3}{2}}}{2}.\nonumber\\
\end{eqnarray}
The coefficients in front of the curvature polynomials take care 
of the subtraction of the ghosts and therefore the result for the $\eta$ is for the
`physical gravitino'. 

Evaluating the curvature polynomials just as in the spin $1/2$ case we obtain
\begin{eqnarray}\label{eta32}
\eta_{3/2}&=&-\frac{1}{(2\pi)^2}\frac{275}{240}\int db dy dx dz a_2'(b)a_3'(y)+\frac{450}{7}I_{CS_2}+490I_{CS_1}.
\end{eqnarray}
Here we have dropped the contribution of the index of the spin 3/2 operator since 
it is an integer for the solid torus. Now the pure topological term in  $\eta_{3/2}$
is extracted by removing the Chern-Simon terms. Finally we also substitute 
the possible winding configurations given in (\ref{q1}) and (\ref{q2}) for the 
graviphoton fields $a_2, a_3$.  This reduces (\ref{eta32}) to 
\begin{eqnarray}
\eta_{3/2} &=& -\frac{275}{60} nm , \\ \nonumber
&=& -\frac{35}{60}nm  - 4nm = -\frac{35}{60} \quad {\rm mod}\; 2.
\end{eqnarray}
Therefore up to mod 2 we can write $\eta_{3/2}$ as
\begin{eqnarray}
\eta_{3/2} = -\frac{1}{(2\pi)^2}\frac{35}{480}\int db dy dx dz a_2'(b)a_3'(y).
\end{eqnarray}
The effective action which reproduces this phase shift under
$a_1 \rightarrow a_1 + 2$ is given by 
\begin{equation}
S_{eff} =  - \frac{ i 35 \pi }{480 (2\pi )^3} \int db dy  dx da dz  a_1(a) a_2'(b)a_3'(y).
\end{equation}
Note that this is $5$ times the result obtained for the Weyl fermion in (\ref{6deff}). 
Therefore  on decompactifying the spatial directions 
and extracting out the transport coefficient  for the gravitinos we obtain
\begin{equation}
\lambda_{3 (3/2)}^{(6)} = \frac{35 \pi}{320 \beta^4}.
\end{equation}
The above result coincides with that obtained using  perturbation theory 
at one loop \cite{Chowdhury:2015pba}.  

The general pattern seen for the contribution of the gravitino to the transport 
coefficient in $2d$ dimensions is that its value is  $2d-1$ times that the result for the 
Weyl fermion. It is  remarkable that  mod 2 ambiguity in determining 
the thermal effective action using  global anomalies is 
consistent with this value of the transport coefficient for the gravitino. 

\subsection{Self-dual tensors}

The APS index theorem for self-dual tensors  in $d=6$ is given by
\begin{eqnarray}
\frac{\sigma_S(B)}{8} &=& \frac{-1}{6!}\int_B \left(\frac{28 p_2(R)}{2}-\frac{16}{8}
p_1( R)^2\right)-\frac{1}{6!\times (2\pi)^4} \int_\Sigma \left(\frac{28}{8}\textrm{Tr}(\theta \wedge R\wedge R\wedge R)\right.\nonumber\\
&&\left. -\frac{28}{16}\textrm{Tr}(\theta \wedge R)\textrm{Tr}(R\wedge R)+\frac{16}{32}\textrm{Tr}(\theta \wedge R)\textrm{Tr}(R\wedge R)\right)-\frac{\eta_A}{2}.\nonumber\\
\end{eqnarray}
Going through the same steps of evaluating the curvature polynomial 
and using the fact that the Hirzebruch index for a solid torus vanishes we obtain 
\begin{eqnarray}\label{selfd6}
\eta_{A}&=& \frac{-16}{240 (2\pi)^2 } \int db dy da dz a_2'(b)a_3'(y)+ 56 I_{CS1}+ \frac{32}{7} I_{CS2}.
\end{eqnarray}
For the refined  global anomaly expression  of \cite{Monnier:2011rk}  the term $\sigma(B)/8$ is 
replaced by $\lambda\wedge \lambda$ where $\lambda$ is a $4$-form. 
Now for the solid torus $B$ which is a disc  times a torus, $D_2 \times T^6$. 
The relative cohomology of the disc has a unique generator 
of degree $2$. This ensures that the intersection pairing 
in degree $4$ of the relative cohomology of $D_2 \times T^6$ vanishes. 
For this situation we  can take $\lambda =0$ \footnote{We again 
thank Samuel Monnier for explaining this to us.}. For the refined global anomaly
formula again there is no need to make an assumption regarding the
Hirzebruch index of $B$ and we obtain the same result as in (\ref{selfd6}). 
Extracting  out the topological term by dropping the Chern-Simons contribution 
we obtain 
\begin{equation}
\eta_A = \frac{-16}{240 (2\pi)^2 } \int db dy dx dz a_2'(b)a_3'(y).
\end{equation}
The thermal effective action  which reproduces this phase shift under $a_1 \rightarrow a_1 + 2$
is given by 
\begin{equation}
S_{{\rm eff}}  = \frac{-i 16\pi }{480 (2\pi)^3 } \int db dy  dx da dz a_1(a)  a_2'(b)a_3'(y).
\end{equation}
Now going through the same steps of decompactifying  the spatial directions and 
extracting out the transport coefficient we obtain the following result 
for self-dual tensors
\begin{equation} \label{anselft}
\lambda_{3(S)}^{(6)} =  \frac{16 \pi }{320 \beta^4}  = \frac{\pi}{20 \beta^4}.
\end{equation}
In the next section we will verify this result by an explicit perturbative 
calculation for the self dual tensors.

\section{ Transport for self dual tensors in $d=6$ at one loop }
 \label{gs6d}
 
Self dual tensors in $4k+2$ dimensions have no Lorentz invariant action though
 they have Lorentz covariant equations of motion. 
 Pure gravitational anomalies exhibited by these theories 
 were studied in perturbation theory by \cite{AlvarezGaume:1983ig}. 
 They proposed Feynman rules and the propagator for these fields
 by which gravitational anomalies in these theories were evaluated. 
 In this section we use these rules at finite temperature to 
 evaluate the transport coefficient $\lambda_3^{(6)}$ in $d=6$. 
 The field strength of  the self dual anti symmetric tensor is defined by 
 \begin{equation}
  F_{\mu_1\mu_2\mu_3} = \partial_{\mu_1} A_{\mu_2 \mu_3} + ( {\rm cyclic\;permutations}), 
 \end{equation}
 where $A_{\mu_1\mu_2}$  is the  2nd rank anti-symmetric gauge potential.  
 The self dual condition  in Euclidean space is given by 
\begin{eqnarray}\label{selfdual}
F^{\mu_1\mu_2\mu_3} &=&
\frac{i}{3! \sqrt{g}} \epsilon^{\mu_1\mu_2\mu_3\nu_1\nu_2\nu_3} F_{\nu_1\nu_2\nu_3}
\equiv i \tilde F^{\mu\nu} ,
\end{eqnarray}
where the  orientation is chosen by setting
\begin{eqnarray}
\epsilon^{\tau a z x y b}=1.
\end{eqnarray}

Let us consider the theory of self dual tensors coupled to metric fluctuations. 
We work in Euclidean space with the signature $\eta^{\mu\nu} = {\rm diag} ( -1, -1, -1, -1, -1, -1) $
The Kubo formula for the transport coefficient of interest in Euclidean space  is given by 
\begin{eqnarray}\label{kubo}
\tilde{\lambda}^{(6)}_3 = 
 -\frac{3}{2} \lim_{p_b,k_y \rightarrow 0}
\frac{\langle T^{\tau a}(k+p)T^{\tau x}(-k)T^{\tau z}(-p)\rangle}{ip_b ik_y}. 
\end{eqnarray} 
where $p,k$ are the external momenta. They are chosen such that 
\begin{eqnarray}
p &=& \left\lbrace 0,0,0,0,0,p^b \right\rbrace,  \qquad\qquad 
k =\left\lbrace 0,0,0,0,k^y,0\right\rbrace.
\end{eqnarray} 
The correlator in Minkowski space is related to that in (\ref{kubo}) by 
\begin{equation}\label{acontinuation}
\tilde{\lambda}^{(6)}_3=-i \lambda^{(6)}_3.
\end{equation}
Note that in (\ref{kubo}) we have taken all the external frequencies to zero first. 
The stress tensor for the self dual boson is defined as follows \cite{AlvarezGaume:1983ig}
\footnote{We have fixed the over all sign in the stress tensor by demanding that
it agrees with the  2 dimensional conformal field theory definition of the stress 
tensor for the chiral boson when applied to $d=2$. The reason the sign differs 
from that in  \cite{AlvarezGaume:1983ig} is due to our choice of mostly negative  
signature of space time.}
First consider 
\begin{equation} \label{defcstress}
 T_{\mu\nu} ( F) =-  \frac{1}{2} F_{\mu \alpha \beta} F_{\nu}^{\, \alpha\beta} + \frac{1}{12} 
 g_{\mu\nu} F_{\alpha\beta\gamma} F^{\alpha\beta\gamma}. 
\end{equation}
Now we impose the self dual condition by considering
\begin{equation}\label{self}
 T_{\mu\nu} ( F^+) = T_{\mu\nu} (\frac{1}{2} (  F + i \tilde F ) ). 
\end{equation}
The hydrodynamic correlation function in (\ref{kubo})  includes the following expectation values
\begin{eqnarray} \label{expansion}
  \langle T^{\mu\alpha} T^{\nu\beta}T^{\rho\sigma} \rangle_E &=& 
  \langle T^{\mu\alpha}_{fl} T^{\nu\beta}_{fl}T^{\rho\sigma}_{fl} \rangle_E \\ \nonumber
& & - 2\langle  \frac{\delta T^{\mu \alpha}}{ \sqrt{g} \delta g_{\nu \beta} } T^{\rho \sigma} \rangle_E
  - 2 \langle  \frac{\delta T^{\mu \alpha}}{ \sqrt{g} \delta g_{\rho \beta} } T^{\nu \sigma}\rangle_E\\ \nonumber
& & -2 \langle T^{\mu\alpha} \frac{\delta T^{\nu \beta}}{ \sqrt{g} \delta g_{\rho \sigma} } \rangle_E 
+ 4 \langle \frac{\delta^2 T^{\mu\alpha}}{\sqrt{ g} \delta g_{\nu \alpha} \delta g_{\rho\sigma}} \rangle_E.
\end{eqnarray}
All these expectation values are taken in the Euclidean vacuum. 
The first term on the RHS is the stress tensor evaluated in flat space.  
All the rest of the terms are contact terms which need to be evaluated 
carefully.  Note that in each of the stress tensor insertions  we need to 
impose the self dual condition by using (\ref{self}). 

Before we discuss the contact  terms, we will present the propagator to evaluate these correlators. 
The thermal 2-point function of the gauge invariant fields is given by 
\begin{eqnarray}\label{propagator}
S_B( \omega_n, p) &=& \langle F^{\mu_1\mu_2\mu_3}(\omega_n, p )F^{\nu_1\nu_2\nu_3}( - \omega_{n'},  - p_3 ) \rangle, \nonumber\\
&=&\left(\frac{p^{\mu_1}p^{\nu_1}g^{\mu_2\nu_2}g^{\mu_3\nu_3}}{\omega_n ^2+ p^2 }+
\text{Permutations}\right)  \beta \delta_{n, n'}  (2\pi)^5 \delta^5( p-p_3), 
\end{eqnarray} 
where, the frequencies for the bosons 
in the Euclidean theory are even  multiples of $\pi T$ and are  given by 
\begin{equation} 
  \omega_n    =   2n \pi T , \qquad\qquad  n\in  \mathbb{Z}.
\end{equation}    

Lets now discuss how  we proceed to evaluate each of the terms in 
(\ref{expansion}). 
The first term is obtained by evaluating the Wick contractions of the  flat space stress 
tensor written in momentum space.  To be explicit we write down the $(\tau x)$  component 
of the stress  tensor
\begin{eqnarray}\label{Stresstensor}
T^{\tau x}_{fl} (-k)&=& - \frac{1}{\beta}\Sigma_{\omega_m}\int \frac{d^5 p_2}{(2\pi)^5}
\left\{ F^{\tau ab}(-p_2-k)F^{xab}(p_2)+F^{\tau ay}(-p_2-k)F^{xay}(p_2) \right. \nonumber\\
&&+F^{\tau az}(-p_2-k)F^{xaz}(p_2)+F^{\tau by}(-p_2-k)F^{xby}(p_2)\nonumber\\
&&\left. +F^{\tau bz}(-p_2-k)F^{xbz}(p_2)+F^{\tau yz}(-p_2-k)F^{xyz}(p_2)\right\}.
\end{eqnarray}
Here the dependence on the Matsubara frequency in the integrand is present in $p_2$ 
whose time component is $\omega_m$, which we have not 
been explicit. 
Note that though we need to impose the self dual projection  on the  field 
strength at every insertion of the stress tensor it is sufficient to  work with the self dual 
insertion on one of the insertions of the stress in the Wick contractions
\cite{AlvarezGaume:1983ig}.  The details of all the Wick contractions 
are performed in the appendix  (\ref{wc6d}). 
To evaluate the contact terms in (\ref{expansion})  we expand the 
stress tensor in (\ref{defcstress}) by considering  only 
metric fluctuations $h_{\tau x} (k) $ and $h_{\tau z}(p) $. 
Here we write down an example of the action of the derivative with respect to 
$h_{\tau z}$ on $T^{\tau x}$
\begin{eqnarray}\label{1stexpansionm}
\frac{\delta T^{\tau x}(-k)}{\delta h_{\tau z}(p)} &=&- \sum_{\omega_m} \int\frac{d^5 p_3}{(2\pi)^5}
\left\{ F^{zab}(-p_3-p-k)F^{xab}(p_3) +F^{zay}(-p_3-p-k)F^{xay}(p_3) \right. \nonumber\\
&&\left. \qquad \qquad \qquad +F^{zby}(-p_3-p-k)F^{xby}(p_3) \right\}.
\end{eqnarray}
Here again we have suppressed the dependence of the Matsubara frequency in the time 
component of $p_3$. 
We have to Wick contract the above expression with $T^{\tau a} ( k + p) $ on which the 
self dual projection is inserted. Similar terms are written down for these class 
of contact terms.  In the appendix \ref{wc6d}, it is shown in detail how 
all these contact terms yield vanishing contribution to the transport coefficient. 
Finally we have the last contact term in (\ref{expansion}) resulting from 
two derivatives of the metric on the stress tensor. 
It is shown in the appendix that this term also vanishes. 
In summary we do not have any contribution from the contact terms. 

The analysis of all possible Wick contractions is tedious and has to be done 
very methodically. This is also performed in detail in the appendix \ref{wc6d}. 
After the Wick contractions there are angular integrals over the internal momenta 
to be performed. We perform these integrals using the method developed in 
\cite{Chowdhury:2015pba}. Essentially we take the zero  
external frequency limit first and then 
take the external momenta to zero before the integration. 
The integrands then simplify considerably and the integrals are easily 
performed.   Finally  all the 
finite terms  resulting from the Wick 
contractions in the zero momentum limit of the correlator  (\ref{kubo}) are given in 
 (\ref{finitecontributions}). 
 The end result of this long calculation  yields the following result for the 
 transport coefficient
\begin{eqnarray}
 \lambda^{(6)}_{3(S)}   &=&\frac{\pi T^4}{20} .
\end{eqnarray} 
This agrees with (\ref{anselft}), the result obtained using global anomaly matching.

\section{Conclusions} \label{concl}

We have used  the method of global anomaly matching put forward in 
\cite{Golkar:2015oxw} 
 for theories with chiral gravitinos and self dual tensors
to determine thermal effective actions and therefore parity odd transport 
coefficients. 
For the case of gravitinos, we obtain results for transport coefficients which are consistent
with perturbative calculations of \cite{Chowdhury:2015pba} up to mod 2. 
Our analysis in $d=6$ shows that the second Pontryagin class does not 
contribute to the topological $\eta$ invariant and therefore does not contribute
to transport. This provides a topological explanation for this  observation in the 
replacement rule of \cite{Loganayagam:2011mu}. 
Finally we have  evaluated the transport coefficients of self dual tensors in $d=6$ using
the Feynman rules put forward in \cite{AlvarezGaume:1983ig}. 
As  far as we are aware this is the first instance where the 
Feynman rules proposed by \cite{AlvarezGaume:1983ig} for the self dual tensor has
been used at finite temperature.  It is indeed satisfying that 
the result agrees with the expectation from global anomaly matching 
as well as the `replacement rule' of \cite{Loganayagam:2011mu}.

From the study in this paper it is clear that  these transport coefficients are 
not perturbatively renormalized \footnote{Recently non-renormalization 
of anomalous transport coefficients was studied holographically in 
\cite{Grozdanov:2016ala}.}
since they are related to global anomalies up to 
mod 2.   
However it will be interesting to figure out the reason which can invoked to 
fix the mod 2 ambiguity in the method of global anomalies. 
One thing we have roughly checked that is  this ambiguity persists 
also for gravitinos in $d=10$. 
It is also of interest to note that it is more easy to determined the 
$\eta$ invariant using the replacement rule of 
\cite{Loganayagam:2011mu} that the direct calculation of various 
curvature invariants. 
There are no holographic checks for 
the transport coefficients of 
self dual tensors and gravitinos since these fields  usually do not occur as dynamic
fields  in 
boundary theories. But, it will be interesting to devise other situations 
where contribution to  transport coefficients from these fields can be 
checked.  Lastly, we claim that the $\eta$ invariant calculated by using the index theorem can be
verified by the computation of various correlators in weak coupling regime. This provides
an easier alternative way to compute the $\eta$-invariant upto a factor of mod 2. 
In appendix \ref{summ}
we have evaluated the $\eta$ invariant for the $T^2$ transformation 
for various chiral matter in $d=10$ using the replacement rule of \cite{Loganayagam:2011mu}
\footnote{We thank R. Loganayagam for suggesting us to carry out this exercise.}.
It is easy to repeat this exercise for arbitrary even dimensions.

\acknowledgments

We thank R. Loganayagam for numerous discussions and suggestions which helped 
clarify our understanding of the issues related to this project. 
We thank Samuel Monnier for very useful correspondence which 
helped us  update our  understanding of the global anomaly formula for 
self dual tensors. He also helped us in understanding how the 
refined global anomaly formula applies for the $T$-diffeomorphisim dealt 
with in this paper. 
We also thank Kallol Sen for helping us automate  performing 
 Wick contractions in Mathematica.

\appendix

\section{Curvature integrals and Chern-Simons terms}\label{Appendix}

It is useful to list  both the vielbeins used in evaluating  the difference between  the 
spin connections of the bulk and the product metric which is defined by 
\begin{equation}
 \theta = \omega_{\mu b}^a ( B) - \omega_{\mu b}^a ( P). 
\end{equation}

\subsection*{2d vielbeins}
The  vielbein which was used for evaluating the spin connection $\omega_{\mu b}^a ( B)$ for the 
metric in (\ref{metricbulk})  are given by 
\begin{eqnarray}
&  & e^{\hat{r}}_r(B) = 1, \qquad e^{\hat{y}}_y(B) = 1,\\ \nonumber
&  & e^{\hat{t}}_t(B) = f(r), \qquad e^{\hat{t}}_x(B) =  f(r) \left(a(x)+2 y\right),\qquad 
e^{\hat{x}}_x(B) = 1.
\end{eqnarray}
The  vielbein used of evaluating the spin connection $\omega_{\mu b}^a ( P)$ for the metric
in (\ref{metricboundary}) are given by 
\begin{eqnarray}
& & e^{\hat{r}}_r(P) = 1, \qquad e^{\hat{y}}_y (P) = 1, \\ \nonumber 
& & e^{\hat{t}}_t (P) = f(1), \qquad e^{\hat{t}}_x(P) =  f(1) \left(a(x)+2 y\right),\qquad e^{\hat{x}}_x (P) = 1 .
\end{eqnarray}

\subsection*{6d vielbeins}
To study the global anomalies in 6d we evaluate $\theta_{\mu}^a(B)$  for the 
metric in (\ref{metric6d}) using the following vielbeins
\begin{eqnarray}
& & e^{\hat{r}}_r(B)  = 1, \qquad e^{\hat{y}}_y (B) = 1,
\qquad e^{\hat{x}}_x (B) = 1\qquad e^{\hat{z}}_z(B)  = 1\nonumber\\
& & e^{\hat{c}}_c (B) = 1,\qquad e^{\hat{b}}_b (B) = 1,\qquad e^{\hat{e}}_e(B)  = 1, \qquad
 e^{\hat{t}}_t(B)  = f(r),   \\  \nonumber
 & &  e^{\hat{t}}_x (B) =  f(r) \left(a_1(a)+ 2u\right),\qquad 
e^{\hat{t}}_z(B)  = f(r) a_2(b),\qquad e^{\hat{t}}_b =f(r)a_3(y).
\end{eqnarray}
For the  product metric at the boundary given in (\ref{metric6dp}) the vielbeins are
\begin{eqnarray}
& & e^{\hat{r}}_r(P)  = 1, \qquad e^{\hat{y}}_y(P)  = 1,\qquad e^{\hat{x}}_x (P)= 1
\qquad e^{\hat{z}}_z(P)  = 1,\nonumber\\
& &  e^{\hat{c}}_c (P) = 1,\qquad e^{\hat{b}}_b(P)  = 1,\qquad e^{\hat{e}}_e(P)  = 1, 
\qquad e^{\hat{t}}_t(P)  = f(1),  \\  \nonumber 
 & & e^{\hat{t}}_x (P) =  f(1) \left(a_1(a)+ 2u\right),\qquad 
 e^{\hat{t}}_z(P)  = f(1) a_2(b),\qquad e^{\hat{t}}_b(P)  =f(1)a_3(y). \nonumber
\end{eqnarray}

\subsection*{Curvature integrals}

To evaluate the $\eta$ invariant in $d=6$ we require the following curvature 
integrals of the metric in (\ref{metric6d}) as well as boundary terms 
associated with the metric in (\ref{metric6dp})
\begin{eqnarray}\label{curvint}
& & I_1 = \int_B  \textrm{Tr} R\wedge R \wedge R \wedge R\nonumber\\
&=& \quad \int_B
2 a_2'(b) a_3'(y) f'(r) \left\{ 
f(r)^7 \left[5+5 a_2'(b)^4-a_3'(y)^2+5 a_3'(y)^4-a_2'(b)^2 \left(1+a_3'(y)^2\right)\right]\right. \nonumber\\
&&\quad \left. +24 f'(r)^2 f''(r)\right\},\nonumber\\
&=& \qquad  \int dt da db dx dy dz   \times \nonumber \\ 
& & 4  \left\{ 
a_2'(b) a_3'(y) \left[ \frac{f(1)^8}{8} \left(5+5 a_2'(b)^4-a_3'(y)^2+5 a_3'(y)^4-a_2'(b)^2 
\left(1+a_3'(y)^2\right)\right)+6f'(1)^4-6\right] \right\},\nonumber\\
\nonumber\\
& & I_{\theta 1} = 
\int_\Sigma  \textrm{Tr} \theta \wedge R \wedge R \wedge R ,
=24 \int dt da db dx dy dz f'(1)^4a_2'(b) a_3'(y),\nonumber\\
\nonumber\\
& & I_2= \int \textrm{Tr} R\wedge R  \wedge \textrm{Tr} R \wedge R,\nonumber\\
&=& \quad \int_B  4a_2'(b) a_3'(y) f'(r) 
\left\{f(r)^7 \left[5+5 a_2'(b)^4+11 a_3'(y)^2+5 a_3'(y)^4+11 a_2'(b)^2 
\left(1+a_3'(y)^2\right)\right]\right.\nonumber\\
&&\quad \left.+16 f(r)^3 \left(1+a_2'(b)^2+a_3'(y)^2\right) f'(r)^2+8
\left[f(r)^4 \left(1+a_2'(b)^2+a_3'(y)^2\right)+3 f'(r)^2\right] f''(r) \right\},\nonumber\\
&= & \int dt da db dx dy dz   8a_2'(b) a_3'(y) \times  \nonumber \\
& &  \quad \left\{ \frac{f(1)^8}{8} 
\left[5+5 a_2'(b)^4+11 a_3'(y)^2+5 a_3'(y)^4+11 a_2'(b)^2 
\left(1+a_3'(y)^2\right)\right]\right.\nonumber\\
&&\left.+4 f(1)^4 f'(1)^2 \left(1+a_2'(b)^2+a_3'(y)^2\right)+ 6 f'(1)^4-6\right\},\nonumber\\
\nonumber\\
& & I_{\theta 2}=\int_\Sigma  \textrm{Tr} \theta \wedge R \textrm{Tr} R \wedge R,
 = \int dt da db dx dy dz  \times \nonumber\\
&&\left\{-  \frac{1}{2} a_2'(b) a_3'(y)\left[-32 f(1)^4f'(1)^2\left(1+a_2'(b)^2+a_3'(y)^2\right) -96 f'(1)^4\right]
\right\}, 
\end{eqnarray}
where  the orientation is decided by  $\epsilon^{taxybur}=1$

\subsection*{Chern Simons  terms}

We obtain identities that relate the terms in the anomaly polynomial $p_1^2 and p_2^2$
to exterior derivatives of Chern-Simons terms. 
We then evaluate these Chern-Simons terms for the metric (\ref{metric6dp}). 

 From the definition of the curvature form we have the identity 
\begin{eqnarray}
\textrm{Tr}(R \wedge R\wedge R\wedge R) &=& \textrm{Tr}(d\omega \wedge d\omega \wedge d\omega \wedge d\omega + 4 d\omega \wedge d\omega \wedge d\omega \wedge \omega \wedge \omega\nonumber\\
&& + 4 d\omega \wedge d\omega \wedge \omega \wedge \omega \wedge \omega \wedge \omega + 2d\omega \wedge \omega \wedge \wedge \omega d\omega  \wedge \omega \wedge \omega \wedge \omega \nonumber\\
&& +4d \omega \wedge \omega \wedge \omega \wedge \omega \wedge \omega \wedge \omega \wedge \omega).
\end{eqnarray}     
We observe that each term in the RHS of the above equation  can be written as an exact form using 
 the following identities
\begin{eqnarray}
& & d[ \textrm{Tr}(\omega \wedge d\omega \wedge d\omega \wedge d\omega )]
= \textrm{Tr} d\omega \wedge d\omega \wedge d\omega \wedge d\omega), \nonumber\\
 & & d[\textrm{Tr}(d\omega \wedge \omega \wedge \omega \wedge \omega \wedge \omega \wedge \omega)]
 =2\textrm{Tr}(d\omega \wedge d\omega \wedge \omega \wedge \omega \wedge \omega \wedge \omega)
 + 
 \textrm{Tr}( d\omega \wedge \omega \wedge \omega \wedge d\omega \wedge \omega \wedge \omega),\nonumber\\
& & \frac{2}{5} d[\textrm{Tr}(d\omega \wedge d\omega \wedge \omega \wedge \omega \wedge \omega)
+ \frac{1}{2} \textrm{Tr}(d\omega \wedge \omega \wedge d\omega \wedge \omega \wedge \omega)]
= \textrm{Tr}(d\omega \wedge d\omega \wedge d\omega \wedge \omega \wedge \omega),\nonumber\\
& &\frac{1}{7} d[\textrm{Tr}(\omega \wedge \omega\wedge \omega\wedge \omega\wedge \omega\wedge 
\omega\wedge \omega)] 
= \textrm{Tr} (d\omega \wedge \omega \wedge \omega\wedge \omega\wedge \omega\wedge
\omega\wedge \omega). 
\end{eqnarray}                          
Combining all these identities we obtain 
\begin{eqnarray}
\textrm{Tr}(R \wedge R\wedge R\wedge R) &=&d[\textrm{Tr} (\omega \wedge d\omega\wedge d\omega\wedge d\omega) + \frac{8}{5}\textrm{Tr} (d\omega \wedge d\omega \wedge \omega \wedge \omega \wedge \omega) \nonumber\\
&&+\frac{4}{5}\textrm{Tr} (d\omega \wedge \omega \wedge d\omega  \wedge \omega \wedge \omega) + \frac{4}{7} \textrm{Tr} (\omega \wedge \omega \wedge \omega \wedge \omega \wedge \omega \wedge \omega \wedge \omega) \nonumber\\
&&+ 2\textrm{Tr} (d\omega \wedge \omega\wedge \omega\wedge \omega\wedge \omega\wedge \omega)].
\end{eqnarray}     
Similarly,
\begin{eqnarray}
\textrm{Tr}(R \wedge R) \textrm{Tr} (R \wedge R) &=& d[ \textrm{Tr}(\omega \wedge d\omega +\frac{2}{3} \omega \wedge \omega \wedge\omega)\wedge \textrm{Tr}(R\wedge R)]. 
\end{eqnarray}                   
Finally we can write down the anomaly polynomials as exterior derivatives of 
Chern-Simons terms. 
\begin{eqnarray}
p_1({\cal{R}})^2&=&\frac{1}{4(2\pi)^4} \frac{\textrm{Tr}(R \wedge R)^2}{4} ,\nonumber\\
&=&\frac{1}{(2\pi)^4}
d\left[\textrm{Tr}(\omega \wedge d\omega +\frac{2}{3} \omega \wedge \omega \wedge\omega) \wedge \textrm{Tr}(R\wedge R)\right] .
\end{eqnarray}
\begin{eqnarray}
&&p_2({\cal{R}})=\frac{-1}{(2\pi)^4}\left(\frac{1}{4}\textrm{Tr}(R \wedge R\wedge R\wedge R)-\frac{1}{8}\textrm{Tr}(R \wedge R)\textrm{Tr}( R\wedge R)\right),\nonumber\\
&=& \frac{-1}{4(2\pi)^4}
d\left[  \textrm{Tr}( \omega \wedge d\omega\wedge d\omega\wedge d\omega )+ 
\frac{8}{5}\textrm{Tr}( d\omega \wedge d\omega \wedge \omega \wedge \omega \wedge \omega)+\frac{4}{5}\textrm{Tr} (d\omega \wedge \omega \wedge d\omega  \wedge \omega \wedge \omega)+ \right.\nonumber\\
&&\left. + \frac{4}{7} \textrm{Tr} (\omega \wedge \omega \wedge \omega \wedge \omega \wedge \omega \wedge \omega \wedge \omega ) + 2\textrm{Tr}( d\omega \wedge \omega\wedge \omega\wedge \omega\wedge \omega\wedge \omega )\right.\nonumber\\
&&\left. -\frac{1}{2} \textrm{Tr}(\omega \wedge d\omega +\frac{2}{3} \omega \wedge \omega \wedge\omega ) \wedge \textrm{Tr}(R\wedge R)  \right] . \nonumber\\
\end{eqnarray}                    

\subsection*{Chern Simons integrals}

We will require integrals of the various Chern Simons forms  over the boundary of the 
metric given in (\ref{metric6d}). 
These are given by 
\begin{eqnarray}
\label{csintegrals}
& & \int_\Sigma \textrm{Tr} \omega \wedge d\omega\wedge d\omega\wedge d\omega = -\frac{ 3(2\pi)^4 }{2} \int db dy
  f(1)^8 a_2'(b) a_3'(y) \left[ \left(a_2'(b)^4+a_3'(y)^4\right)+1\right], \nonumber\\
&&
\int_\Sigma\textrm{Tr} d\omega \wedge d\omega \wedge \omega \wedge \omega \wedge \omega 
= -\frac{ (2\pi)^4}{2}  \int db dy 
  f(1)^8 a_2'(b) a_3'(y) \left[ \left(a_2'(b)^4+a_3'(y)^4\right)+1\right],\nonumber\\
&& \int_\Sigma \textrm{Tr} d\omega \wedge \omega \wedge d\omega \wedge \omega \wedge \omega = -\frac{(2\pi)^4}{4} \int dbdy
  f(1)^8 a_2'(b) a_3'(y) \left[  \left(a_2'(b)^4+a_3'(y)^4\right)+1\right], \nonumber\\
&& \int_{\Sigma}\textrm{Tr} d\omega \wedge \omega\wedge \omega\wedge \omega\wedge \omega\wedge \omega = \frac{ (2\pi)^4}{4} \int db dy 
 f(1)^8 a_2'(b) a_3'(y) \left[ a_2'(b)^2 \left( a_3'(y)^2+1\right)+ a_3'(y)^2\right], \nonumber\\
&& \int_{\Sigma}\textrm{Tr} \omega \wedge \omega \wedge \omega \wedge \omega \wedge \omega \wedge \omega \wedge \omega = 0, \\ 
&& \int_{\Sigma}\textrm{Tr}(\omega \wedge d\omega +\frac{2}{3} \omega \wedge \omega \wedge\omega)\textrm{Tr}(R\wedge R) = (2\pi)^4 f(1)^4 \int db dy a_2'(b)a_3'(y)   \times  \nonumber \\
 && \qquad\qquad \qquad \left\{   -16 f'(1)^2 \left(a_2'(b)^2+a_3'(y)^2+1\right)  \right.  \nonumber \\ 
 & &\qquad \qquad \left.  -f(1)^4 \left[ 11 a_2'(b)^2 \left(a_3'(y)^2+1\right)+5 a_2'(b)^4+5 a_3'(y)^4  +11 a_3'(y)^2+5\right] \right\} .\nonumber
\end{eqnarray}
In the last integral it is important to note that the curvature form ${\rm Tr } (R\wedge R)$ is 
 evaluated from the metric  (\ref{metric6d}) but at the boundary. 
\newpage

\section{Correlators of self dual tensors in $d=6$ }\label{wc6d}

In this appendix we present the details involved in evaluating the 
Kubo formula (\ref{kubo}) to obtain the transport coefficient $\lambda^{(6)}_{3(S)}$ 
We first write down the components of the flat space stress tensors we will use explicitly. 
This will facilitate the discussion of the Wick contractions. 
\begin{eqnarray}
T^{\tau a}_{fl} (p+k) &=& \frac{-1}{\beta}\sum_{\omega_m}\int \frac{d^5 p_1}{(2\pi)^5}\left[\frac{1}{2}\left(F^{\tau bx}+iF^{ayz}\right)\left(-p_1+p+k\right)\left(F^{abx}-iF^{\tau yz}\right)(p_1)\right.\nonumber\\
&&\left.+\frac{1}{2}\left(F^{\tau by}-iF^{axz}\right)\left(-p_1+p+k\right)\left(F^{aby}+iF^{\tau xz}\right)(p_1)\right.\nonumber\\
&&\left.+\frac{1}{2}\left(F^{\tau bz}-iF^{ayx}\right)\left(-p_1+p+k\right)\left(F^{abz}+iF^{\tau yx}\right)(p_1)\right] ,\nonumber\\
T^{\tau x}_{fl}(-k)&=&\frac{-1}{\beta}\sum_{\omega_m}\int \frac{d^5 p_2}{(2\pi)^5} \left[ (F^{\tau ab}(-p_2-k)F^{xab}(p_2)+F^{\tau ay}(-p_2-k)F^{xay}(p_2)\right.\nonumber\\
&&\left.+F^{\tau az}(-p_2-k)F^{xaz}(p_2)+F^{\tau by}(-p_2-k)F^{xby}(p_2)\right.\nonumber\\
&&\left.+F^{\tau bz}(-p_2-k)F^{xbz}(p_2)+F^{\tau yz}(-p_2-k)F^{xyz}(p_2)\right], \nonumber\\
T^{\tau z}_{fl} (-p)&=&\frac{-1}{\beta}\sum_{\omega_m}\int \frac{d^5 p_3}{(2\pi)^5} \left[ 
F^{\tau ab}(-p_3-p)F^{zab}(p_3)+F^{\tau ay}(-p_3-p)F^{zay}(p_3) \right.\nonumber\\
&&\left.+F^{\tau ax}(-p_3-p)F^{zax}(p_3)+F^{\tau by}(-p_3-p)F^{zby}(p_3)\right.\nonumber\\
&&\left. +F^{\tau bx}(-p_3-p)F^{zbx}(p_3)+F^{\tau yx}(-p_3-p)F^{zyx}(p_3)\right].\nonumber\\
\end{eqnarray}
We have imposed the self dual projection on $T^{\tau a}$, also note that after summing over 
the indices involved in $F$ we obtain an overall  factor of $1/2$. 
Again in writing down $T^{\tau x}$ and $T^{\tau z}$, the indices in $F$ has been summed over. 

The kinematic configurations is same as the one used in \cite{Chowdhury:2015pba} for $d=6$.
The two external momenta are labelled as $p, k $. The momentum $p$ has non-zero
component only in the $b$ direction, while $k$ has non-zero momentum only in the 
$y$ direction. The one loop  internal momenta is  labelled by $p_3$. To perform the 
integration over $p_3$, we will parametrize its components in terms of angular variables
\begin{eqnarray}
p_3^{b} = |p_3| \cos\phi_1, \quad p_3^{y} =|p_3| \sin\phi_1 \cos\phi_2, \quad
p_3^x = |p_3| \sin\phi_1 \sin\phi_2 \cos\phi_3, \\ \nonumber
 p_3^z = \sin\phi_1 \sin\phi_2 \sin \phi_3 \cos\phi_4, \quad 
 p_3^a = \sin\phi_1 \sin\phi_2\sin\phi_3 \sin\phi_4.
\end{eqnarray}
With this kinematic configuration,  we  write down the expression for the energies 
which will occur in the expression for the propagators. 
\begin{eqnarray}\label{energies}
E_{p_3+p}&=&( {|p_3|}^2+{|p|}^2+2|p||p_3| \cos \phi_1)^{\frac{1}{2}},  \\ \nonumber
E_{p3+k}&=& ( {|p_3|}^2+{|k|}^2+2|k||p_3|\sin \phi_1 \cos \phi_2 )^{\frac{1}{2} } ,\\  \nonumber
E_{p3-k}&=& ( {|p_3|}^2+{|k|}^2-2|k||p_3|\sin \phi_1 \cos \phi_2 )^{\frac{1}{2} }, \\ \nonumber
E_{p_3+p+k} &=&(
{|p_3|}^2+{|k|}^2+{|p|}^2+2|k||p_3|\sin \phi_1 \cos \phi_2 + 2|p||p_3|\cos \phi_1 )^{\frac{1}{2}}.\nonumber
\end{eqnarray}
 In order to keep track of the  Wick contractions, we adopt the following convention. 
We will denote the $i ^{\rm th}$  term in the expression for the stress tensor as, 
$T^{\mu \nu}_i$. For example,
\begin{eqnarray}
T^{\tau x}_2 &=& \frac{-1}{\beta} \sum_{\omega_n} \int \frac{d^5 p_2}{(2\pi)^5}F^{\tau ay}(-p_2-k)F^{xay}(p_2).                      
\end{eqnarray}                                      
From \eqref{Stresstensor} and 
\eqref{kubo}, we see that the general Wick contraction structure looks like,
\begin{eqnarray}
I=T^{\tau a}_i T^{\tau x}_j T^{\tau z}_k,
\end{eqnarray}
where $i,j,k$ denote the respective terms in the expression for the stress tensor.

\subsection*{\bf Contractions with odd number of $\tau$'s vanish}

There is one important simplification when we consider the Wick contractions before
a detailed evaluation.  
Note that the Matsubara sums run from negative infinity to positive infinity.
 Applying \eqref{propagator} to such a generic contraction, it is easy to see that the 
 Matsubara sum gives non zero contribution only when the terms involved in the Wick contraction have even number of $\tau$ indices.
  Let us illustrate this claim with the help of an example,
\begin{eqnarray}
\langle T^{\tau a}_1 T^{\tau x}_3 T^{\tau z}_1 \rangle &=& \frac{-1}{\beta} \sum_{\omega^1_n} \int \frac{d^5 p_1}{(2\pi)^5}\frac{-1}{\beta} \sum_{\omega^2_n} \int \frac{d^5 p_2}{(2\pi)^5}\frac{-1}{\beta} \sum_{\omega^3_n} \int \frac{d^5 p_3}{(2\pi)^5}\langle \frac{1}{2}F^{\tau bx}(-p_1+p+k)F^{\tau yz}(p_1) \nonumber\\
&& F^{\tau az}(-p_2-k)F^{xaz}(p_2) F^{\tau ab}(-p_3-p)F^{zab}(p_3) \rangle, \nonumber\\
\nonumber \\
&=&\frac{-1}{\beta} \sum_{\omega^3_n} \int \frac{d^5 p_3}{(2\pi)^5}\langle F^{abx}(p_3-k)F^{azx}(-p_3+k)\rangle\langle F^{\tau a z}(p_3) F^{azb}(-p_3)\rangle\nonumber\\
&&\langle F^{\tau b x}(p_3+p) F^{\tau b a}(-p_3-p)\rangle,  \nonumber\\ \nonumber \\
&=&\frac{-1}{\beta} \sum_{\omega^3_n} \int \frac{d^5 p_3}{(2\pi)^5}\frac{-i}{2}\frac{(p_3^b)^2 p_3^z p_3^x p_3^a (i\omega_n^3)}{((i\omega_n^3)^2-E_{p_3-k}^2)((i\omega_n^3)^2-E_{p_3}^2)((i\omega_n^3)^2-E_{p_3+p}^2)}.
\end{eqnarray}
The sum over the Matsubara frequencies  in the above expression is from negative infinity to positive infinity.  Note that the sum runs over an odd function of $\omega^3$.  Therefore the 
result vanishes.

Therefore using the observation in the previous paragraph we can conclude that 
 the terms in the stress tensor $T^{\tau a}$ that result 
in non-zero wick contractions are,
\begin{eqnarray}
T^{\tau a}(p+k) &=& \frac{-1}{\beta}\sum_{\omega_m}\int \frac{d^5 p_1}{(2\pi)^5}\left(-\frac{i}{2}F^{\tau bx}(-p_1+p+k)F^{\tau yz}(p_1)+\frac{i}{2}F^{ayz}(-p_1+p+k)F^{abx}(p_1)\right.\nonumber\\
&&\left.+\frac{i}{2}F^{\tau by}(-p_1+p+k)F^{\tau xz}(p_1)-\frac{i}{2}F^{axz}(-p_1+p+k)F^{aby}(p_1)\right.\nonumber\\
&&\left.+\frac{i}{2}F^{\tau bz}(-p_1+p+k)F^{\tau yx}(p_1)-\frac{i}{2}F^{ayx}(-p_1+p+k)F^{abz}(p_1)\right).\nonumber\\
\end{eqnarray}
This is because the expansion of the stress tensor $T^{\tau x}$ and $T^{\tau z}$ contains
a single $\tau$ in each of its terms.

The resulting Wick contractions even after this simplification, are numerous.  We 
have developed a Mathematica code  to perform the Wick contractions. 
These Wick contractions can be broadly classified into two types depending on the denominators. We illustrate this fact with the following examples.
\begin{eqnarray}\label{sampleCA}
\langle T^{\tau a}_1 T^{\tau x}_1 T^{\tau z}_2\rangle &=& \frac{-1}{\beta} \sum_{\omega^1_n} \int \frac{d^5 p_1}{(2\pi)^5}\frac{1}{\beta} \sum_{\omega^2_n} \int \frac{d^5 p_2}{(2\pi)^5}\frac{1}{\beta} \sum_{\omega^3_n} \int \frac{d^5 p_3}{(2\pi)^5}\langle-\frac{i}{2}F^{\tau bx}(-p_1+p+k)F^{\tau yz}(p_1) \nonumber\\
&& \qquad\qquad 
 \times F^{\tau ab}(-p_2-k)F^{xab}(p_2)F^{\tau ay}(-p_3-p)F^{zay}(p_3)\rangle,\nonumber\\
&=&\frac{1}{\beta}\sum_{\omega^3_n} \int \frac{d^5 p_3}{(2\pi)^5}
\frac{i}{2}\frac{\{-p_3\}^a \{-p_3\}^\tau \{p+p_3\}^b \{p+p_3\}^y \{k+p+p_3\}^a \{k+p+p_3\}^\tau}{\left(i\omega^2-E_{(p_3)^2}\right) \left(i\omega^2-E_{(p+p_3)^2}\right) \left(i\omega^2-E_{(k+p+p_3)^2}\right)}.\nonumber\\ 
\end{eqnarray}
\begin{eqnarray}
\label{sampleCB}
\langle T^{\tau a}_5 T^{\tau x}_1 T^{\tau z}_5 \rangle &=&\frac{-1}{\beta} \sum_{\omega^1_n} \int \frac{d^5 p_1}{(2\pi)^5}\frac{1}{\beta} \sum_{\omega^2_n} \int \frac{d^5 p_2}{(2\pi)^5}\frac{1}{\beta} \sum_{\omega^3_n} \int \frac{d^5 p_3}{(2\pi)^5} \langle \frac{i}{2}F^{\tau bz}(-p_1+p+k)F^{\tau yx}(p_1)\nonumber\\
&& \qquad \qquad \times 
F^{\tau ab}(-p_2-k)F^{xab}(p_2)F^{\tau bx}(-p_3-p)F^{zbx}(p_3)\rangle, \nonumber\\
&=&\frac{1}{\beta}\sum_{\omega^3_n} \int \frac{d^5 p_3}{(2\pi)^5}
\frac{i}{2}\frac{\{-p_3\}^a \{-p_3\}^z \{k-p_3\}^a \{p+p_3\}^b \{k-p_3\}^z \{p+p_3\}^y}{\left(i\omega^2-E_{(-p_3)^2}\right) \left(i\omega^2-E_{(k-p_3)^2}\right) \left(i\omega^2-E_{(p+p_3)^2}\right)}.\nonumber\\
\end{eqnarray}

\newpage
\subsection*{Result of the Wick contractions}

All the Wick contractions fall into two classes depending on the denominator. 
They are given by  terms of the kind 
\begin{eqnarray}\label{cA}
W_A &=& \frac{1}{\beta}\sum_{\omega^3_n} \int \frac{d^5 p_3}{(2\pi)^5} c_A, \nonumber \\
c_A&=& \{ [-i (k^y)^2 ((p_3^a)^2 p_3^b p_3^y+p_3^b (p_3^\tau)^2 p_3^y+p_3^b (p_3^x)^2 p_3^y+p_3^b p_3^y (p_3^z)^2\nonumber\\
&&+(p_3^b)^3 p_3^y+p_3^b (p_3^y)^3)-i p_3^b k^y ((p_3^a)^2+(p_3^b)^2-(p_3^\tau)^2+(p_3^x)^2+(p_3^y)^2+(p_3^z)^2)\nonumber\\
&&((p_3^a)^2+(p_3^b)^2+(p_3^\tau)^2+(p_3^x)^2+(p_3^y)^2+(p_3^z)^2)]+[-i k^y ((p_3^a)^2 (4 (p_3^b)^2-3 (p_3^\tau)^2\nonumber\\
&&+2 ((p_3^x)^2+(p_3^z)^2))+(p_3^a)^4+(p_3^b)^2 (-(p_3^\tau)^2+4 (p_3^x)^2+2 (p_3^y)^2+4 (p_3^z)^2)+3 (p_3^b)^4\nonumber\\
&&-((p_3^x)^2+(p_3^y)^2+(p_3^z)^2) (3 (p_3^\tau)^2-(p_3^x)^2+(p_3^y)^2-(p_3^z)^2))+i p_3^y ((p_3^a)^2+(p_3^b)^2\nonumber\\
&&+(p_3^x)^2+(p_3^y)^2+(p_3^z)^2) ((p_3^a)^2+(p_3^b)^2+(p_3^\tau)^2+(p_3^x)^2+(p_3^y)^2+(p_3^z)^2)\nonumber\\
&&-i (k^y)^2 ((p_3^b)^2 p_3^y-(p_3^\tau)^2 p_3^y+p_3^y (p_3^z)^2)]p^b+
[k^y (-3 i (p_3^a)^2 p_3^b+i p_3^b (p_3^\tau)^2\nonumber\\
&&-2 i p_3^b (p_3^x)^2-3 i p_3^b (p_3^z)^2-3 i (p_3^b)^3)+(3 i (p_3^a)^2 p_3^b p_3^y+i p_3^b (p_3^\tau)^2 p_3^y\nonumber\\
&&+3 i p_3^b (p_3^x)^2 p_3^y+3 i p_3^b p_3^y (p_3^z)^2+3 i (p_3^b)^3 p_3^y+3 i p_3^b (p_3^y)^3)](p^b)^2\nonumber\\
&&+[k^y (-i (p_3^a)^2-i (p_3^b)^2-i (p_3^z)^2)+(i (p_3^a)^2 p_3^y+3 i (p_3^b)^2 p_3^y+i (p_3^x)^2 p_3^y\nonumber\\
&&+i p_3^y (p_3^z)^2+i (p_3^y)^3)](p^b)^3+i (p^b)^4 p_3^b p_3^y \}\nonumber\\
&&\times \frac{-1}{2\left(i\omega^2-E_{-p_3}^2\right) \left(i\omega^2-E_{p+p_3}^2\right) \left(i\omega^2-E_{k+p+p_3}^2\right)}.
\end{eqnarray}
\begin{eqnarray}\label{cB}
W_B &=& \frac{1}{\beta}\sum_{\omega^3_n} \int \frac{d^5 p_3}{(2\pi)^5} c_B, \nonumber \\
c_B&=& \{\left[i (k^y)^2 ((p_3^a)^2 p_3^b p_3^y+p_3^b (p_3^\tau)^2 p_3^y+p_3^b (p_3^x)^2 p_3^y+p_3^b p_3^y (p_3^z)^2+(p_3^b)^3 p_3^y \right.\nonumber\\
&&\left. +p_3^b (p_3^y)^3)-i p_3^b k^y ((p_3^a)^2+(p_3^b)^2-(p_3^\tau)^2+(p_3^x)^2+(p_3^y)^2+(p_3^z)^2) ((p_3^a)^2+(p_3^b)^2\right. \nonumber\\
&&\left. +(p_3^\tau)^2+(p_3^x)^2+(p_3^y)^2+(p_3^z)^2)\right]+\left[-i k^y \left(3 (p_3^a)^2 (p_3^b)^2-3 (p_3^a)^2 (p_3^\tau)^2+2 (p_3^a)^2 (p_3^x)^2\right.\right.\nonumber\\
&&\left.\left.  +3 (p_3^a)^2 (p_3^y)^2+2 (p_3^a)^2 (p_3^z)^2+(p_3^a)^4-2 (p_3^b)^2 (p_3^\tau)^2+3 (p_3^b)^2 (p_3^x)^2+4 (p_3^b)^2 (p_3^y)^2\right.\right. \nonumber\\
&&\left.\left.  +3 (p_3^b)^2 (p_3^z)^2+2 (p_3^b)^4-3 (p_3^\tau)^2 (p_3^x)^2-3 (p_3^\tau)^2 (p_3^z)^2+3 (p_3^x)^2 (p_3^y)^2\right.\right. \nonumber\\
&&\left.\left.  +2 (p_3^x)^2 (p_3^z)^2+(p_3^x)^4+3 (p_3^y)^2 (p_3^z)^2+2 (p_3^y)^4+(p_3^z)^4\right)+i (k^y)^2 \left((p_3^a)^2 p_3^y\right.\right. \nonumber\\
&&\left.\left.+2 (p_3^b)^2 p_3^y+(p_3^x)^2 p_3^y+2 p_3^y (p_3^z)^2+(p_3^y)^3\right)+i p_3^y \left((p_3^a)^2 \right.\right. \nonumber\\
&&\left.\left.  +(p_3^b)^2+(p_3^x)^2+(p_3^y)^2+(p_3^z)^2) ((p_3^a)^2+(p_3^b)^2+(p_3^\tau)^2+(p_3^x)^2+(p_3^y)^2+(p_3^z)^2\right)\right]p^b \nonumber\\
&& +\left[-i k^y ((p_3^a)^2 p_3^b-p_3^b (p_3^\tau)^2+2 p_3^b (p_3^x)^2+2 p_3^b (p_3^y)^2+p_3^b (p_3^z)^2+(p_3^b)^3\right. \nonumber\\
&& \left. +i p_3^b p_3^y ((p_3^a)^2+(p_3^b)^2+(p_3^\tau)^2+(p_3^x)^2+(p_3^y)^2+(p_3^z)^2)+i p_3^b (k^y)^2 p_3^y\right](p^b)^2\}\nonumber\\
&&\times \frac{-1}{2\left(i\omega^2-E_{k-p_3}^2\right) \left(i\omega^2-E_{-p_3}^2\right) \left(i\omega^2-E_{p+p_3}^2\right)}. 
\end{eqnarray}
From the analysis it is evident that Wick contractions in 
\eqref{sampleCA} belongs to class \eqref{cA} while the contractions in 
 \eqref{sampleCB} belongs to Class \eqref{cB}

\subsection*{Contractions organised as powers of the external momenta}

We will show that the contact terms in \eqref{expansion} result in vanishing contributions
and therefore the entire contribution to $\lambda^{(6)}_{3(S)}$ arise from the 
Wick contractions of the first term in  \eqref{expansion}. 
We now systematically proceed to evaluate the terms in \eqref{cA} and \eqref{cB} to  obtain the contributions to the transport coefficient. 
The terms in \eqref{cA} and \eqref{cB} can be arranged in polynomials in $p,k$ according 
to the powers of the external momenta in the numerator of these expressions. 

Examining the the terms in \eqref{cA} and \eqref{cB} can be arranged according to 
decreasing powers of the external momenta in the numerator as follows. 
The orders in external momenta are also indicated alongside the terms
\begin{eqnarray}
M_1&=&\frac{-i(p^b)^4 p_3^b p_3^y}{2\left(i\omega^2-E_{-p_3}^2\right) 
\left(i\omega^2-E_{p+p_3}^2\right) \left(i\omega^2-E_{k+p+p_3}^2\right)},
\qquad \qquad \qquad  O(p^4), \nonumber\\
\nonumber\\
M_2 &=&- \frac{ (-i (p_3^a)^2-i (p_3^b)^2-i (p_3^z)^2)k^y(p^b)^3 }
{2\left(i\omega^2-E_{-p_3}^2\right) \left(i\omega^2-E_{p+p_3}^2\right)
\left(i\omega^2-E_{k+p+p_3}^2\right)},  \qquad\qquad O(p^3k), \nonumber\\
\nonumber\\
M_3&=&\frac{-i p_3^b p_3^y (k^y)^2{p^b}^2}
{2\left(i\omega^2-E_{k-p_3}^2\right) \left(i\omega^2-E_{-p_3}^2\right) 
\left(i\omega^2-E_{p+p_3}^2\right)},  \qquad\qquad\qquad O(p^2k^2),\nonumber\\
\nonumber\\
M_4&=&-\frac{(i (p_3^a)^2 p_3^y+3 i (p_3^b)^2 p_3^y+i (p_3^x)^2 p_3^y+i (p_3^y)^3
+i p_3^y (p_3^z)^2)}{2(i\omega^2-E_{-p_3}^2) (i\omega^2-E_{p+p_3}^2) (i\omega^2-E_{k+p+p_3}^2)}
{(p^b)}^3, \qquad\quad O(p^3),\nonumber\\
\nonumber\\
M_5&=&-\frac{(-3 i (p_3^a)^2 p_3^b-3 i (p_3^b)^3+
i p_3^b (p_3^\tau)^2-2 i p_3^b (p_3^x)^2-3 i p_3^b (p_3^z)^2)}
{2(i\omega^2-E_{-p_3}^2) (i\omega^2-E_{p+p_3}^2) (i\omega^2-E_{k+p+p_3}^2)}
k^y {(p^b)}^2\nonumber\\
&&+\frac{i ((p_3^a)^2 p_3^b+(p_3^b)^3-p_3^b (p_3^\tau)^2
+2 p_3^b (p_3^x)^2+2 p_3^b (p_3^y)^2+p_3^b (p_3^z)^2) }
{2\left(i\omega^2-E_{k-p_3}^2\right) \left(i\omega^2-E_{-p_3}^2\right) 
\left(i\omega^2-E_{p+p_3}^2\right)}k^y {(p^b)}^2,  \qquad\quad O(p^2k),\nonumber\\
\nonumber\\
M_6&=&\frac{i ((p_3^b)^2 p_3^y-(p_3^\tau)^2 p_3^y+p_3^y (p_3^z)^2)}
{2(i\omega^2-E_{-p_3}^2) (i\omega^2-E_{p+p_3}^2) 
(i\omega^2-E_{k+p+p_3}^2)}{(k^y)}^2p^b\nonumber\\
 &&+\frac{i ((p_3^a)^2 p_3^y+2 (p_3^b)^2 p_3^y
 +(p_3^x)^2 p_3^y+(p_3^y)^3+2 p_3^y (p_3^z)^2)}
 {2\left(i\omega^2-E_{k-p_3}^2\right) \left(i\omega^2-E_{-p_3}^2\right) 
 \left(i\omega^2-E_{p+p_3}^2\right)} (k^y)^2p^b,  \qquad \qquad O(pk^2),\nonumber\\
 \nonumber\\
 \nonumber\\
 M_7&=&-\frac{(3 i (p_3^a)^2 p_3^b p_3^y+3 i (p_3^b)^3 p_3^y+i p_3^b (p_3^\tau)^2 p_3^y
 +3 i p_3^b (p_3^x)^2 p_3^y+3 i p_3^b (p_3^y)^3+3 i p_3^b p_3^y (p_3^z)^2)}
 {2(i\omega^2-E_{-p_3}^2) (i\omega^2-E_{p+p_3}^2) 
 (i\omega^2-E_{k+p+p_3}^2)}{(p^b)}^2\nonumber\\
&&-\frac{i p_3^b p_3^y ((p_3^a)^2+(p_3^b)^2+
(p_3^\tau)^2+(p_3^x)^2+(p_3^y)^2+(p_3^z)^2)}
{2\left(i\omega^2-E_{k-p_3}^2\right) 
\left(i\omega^2-E_{-p_3}^2\right) \left(i\omega^2-E_{p+p_3}^2\right)}
{(p^b)}^2, \qquad\qquad O(p^2),\nonumber\\
\nonumber\\
\end{eqnarray}
\begin{eqnarray}
M_{8}^A&=&i [(p_3^a)^4+3 (p_3^b)^4-(3 (p_3^\tau)^2-(p_3^x)^2+(p_3^y)^2-(p_3^z)^2) ((p_3^x)^2+(p_3^y)^2+(p_3^z)^2)+(p_3^b)^2\nonumber\\
&&\times(-(p_3^\tau)^2+4 (p_3^x)^2+2 (p_3^y)^2+4 (p_3^z)^2)+(p_3^a)^2 (4 (p_3^b)^2-3 (p_3^\tau)^2+2 ((p_3^x)^2+(p_3^z)^2)))]k^yp^b\nonumber\\
&&\times \frac{1}{2(i\omega^2-E_{-p_3}^2) (i\omega^2-E_{p+p_3}^2) 
(i\omega^2-E_{k+p+p_3}^2)},  \qquad \qquad \qquad\qquad O(pk), \nonumber\\
\nonumber\\
M_{8}^B&=&i ((p_3^a)^4+3 (p_3^a)^2 (p_3^b)^2+2 (p_3^b)^4-3 (p_3^a)^2 (p_3^\tau)^2-2 (p_3^b)^2 (p_3^\tau)^2+2 (p_3^a)^2 (p_3^x)^2\nonumber\\
&&+3 (p_3^b)^2 (p_3^x)^2-3 (p_3^\tau)^2 (p_3^x)^2+(p_3^x)^4+3 (p_3^a)^2 (p_3^y)^2+4 (p_3^b)^2 (p_3^y)^2+3 (p_3^x)^2 (p_3^y)^2+2 (p_3^y)^4\nonumber\\
&&+2 (p_3^a)^2 (p_3^z)^2+3 (p_3^b)^2 (p_3^z)^2-3 (p_3^\tau)^2 (p_3^z)^2+2 (p_3^x)^2 (p_3^z)^2+3 (p_3^y)^2 (p_3^z)^2+(p_3^z)^4)p^bk^y \nonumber\\
&&\times \frac{1}{2\left(i\omega^2-E_{k-p_3}^2\right) 
\left(i\omega^2-E_{-p_3}^2\right) \left(i\omega^2-E_{p+p_3}^2\right)}, 
\qquad \qquad \qquad \qquad O(pk), \nonumber\\
\nonumber\\
\nonumber\\
M_9&=& \frac{i ((p_3^a)^2 p_3^b p_3^y+(p_3^b)^3 p_3^y+p_3^b (p_3^\tau)^2 p_3^y+p_3^b (p_3^x)^2 p_3^y+p_3^b (p_3^y)^3+p_3^b p_3^y (p_3^z)^2){k^y}^2}{2(i\omega^2-E_{-p_3}^2) (i\omega^2-E_{p+p_3}^2) (i\omega^2-E_{k+p+p_3}^2)}\nonumber\\
&&-\frac{{k^y}^2i ((p_3^a)^2 p_3^b p_3^y+(p_3^b)^3 p_3^y+p_3^b (p_3^\tau)^2 
p_3^y+p_3^b (p_3^x)^2 p_3^y+p_3^b (p_3^y)^3+p_3^b p_3^y (p_3^z)^2)}
{2\left(i\omega^2-E_{k-p_3}^2\right) \left(i\omega^2-E_{-p_3}^2\right) 
\left(i\omega^2-E_{p+p_3}^2\right)},  \nonumber\\
& & \qquad\qquad\qquad \qquad \qquad
\qquad\qquad\qquad \qquad \qquad\qquad\qquad\qquad \qquad \qquad O(k^2), 
\nonumber\\
M_{10}&=&i p_3^b [(p_3^a)^2+(p_3^b)^2-(p_3^\tau)^2+(p_3^x)^2+(p_3^y)^2+(p_3^z)^2]
[(p_3^a)^2+(p_3^b)^2+(p_3^\tau)^2+(p_3^x)^2\nonumber\\
&&+(p_3^y)^2+(p_3^z)^2]\frac{1}{2(i\omega^2-E_{-p_3}^2) (i\omega^2-E_{p+p_3}^2) (i\omega^2-E_{k+p+p_3}^2)}k^y \nonumber\\
&&+i p_3^b [(p_3^a)^2+(p_3^b)^2-(p_3^\tau)^2+(p_3^x)^2+(p_3^y)^2+(p_3^z)^2] [(p_3^a)^2+(p_3^b)^2+(p_3^\tau)^2+(p_3^x)^2\nonumber\\
&&+(p_3^y)^2+(p_3^z)^2]\frac{1}
{2\left(i\omega^2-E_{k-p_3}^2\right) \left(i\omega^2-E_{-p_3}^2\right) 
\left(i\omega^2-E_{p+p_3}^2\right)}k^y,    \qquad\qquad  O(k), \nonumber\\
\nonumber\\
\nonumber\\
M_{11} &=& -\frac{i p_3^y ((p_3^a)^2+(p_3^b)^2+(p_3^x)^2+(p_3^y)^2+(p_3^z)^2) ((p_3^a)^2+(p_3^b)^2+(p_3^\tau)^2+(p_3^x)^2+(p_3^y)^2+(p_3^z)^2)}{2(i\omega^2-E_{-p_3}^2) (i\omega^2-E_{p+p_3}^2) (i\omega^2-E_{k+p+p_3}^2)}p^b\nonumber\\
&&-\frac{i p_3^y ((p_3^a)^2+(p_3^b)^2+(p_3^x)^2+(p_3^y)^2+
(p_3^z)^2) ((p_3^a)^2+(p_3^b)^2+(p_3^\tau)^2+(p_3^x)^2+(p_3^y)^2+(p_3^z)^2)}
{2\left(i\omega^2-E_{k-p_3}^2\right) \left(i\omega^2-E_{-p_3}^2\right) 
\left(i\omega^2-E_{p+p_3}^2\right)}p^b, \nonumber\\
&& \qquad\qquad\qquad\qquad  \qquad\qquad\qquad\qquad 
\qquad\qquad\qquad  \qquad\qquad\qquad\qquad O(p),\nonumber\\
\end{eqnarray}

\subsection*{Terms which do not contribute}

We now show that all  contributions from  $M_1$ to $M_7$ vanish on dividing by 
$p^b k^y$, summing over the Matsubara frequencies, performing the angular integrals
 and then taking the zero momentum limit.

\subsubsection*{\small{Vanishing of $M_1$}}

\begin{eqnarray}
I_{M_1}&=&\frac{-1}{p^bk^y}\sum_{\omega_m}\int \frac{d^5p_3}{(2\pi)^5}\frac{i(p^b)^4 p_3^b p_3^y}{2\left(i\omega^2-E_{-p_3}^2\right) \left(i\omega^2-E_{p+p_3}^2\right) \left(i\omega^2-E_{k+p+p_3}^2\right)},\nonumber\\
\nonumber\\
&=&\int \frac{d^5p_3}{(2\pi)^5}\frac{-i(p^b)^3 p_3^b p_3^y}{2k^y}{\cal M}.
\end{eqnarray} 
where we  have performed the Matsubara sum using 
 \begin{eqnarray}\label{ms6d1}
 && {\cal M} \equiv    
\frac{1}{\beta}\sum_{ m}\frac{1}{\left(i\omega_m^2-E_{-p_3}^2\right) \left(i\omega_m^2-E_{p+p_3}^2\right) \left(i\omega_m^2-E_{k+p+p_3}^2\right)},\\ \nonumber
&&=\frac{-b(E_{p3+p+k})}{E_{p3+p+k}(E_{p_3+p+k}^2-E_{p_3}^2)(E_{p_3+p+k}^2-E_{p_3+p}^2)}+
\frac{-b(E_{p_3+p})}{E_{p_3+p}(E_{p_3+p}^2-E_{p_3+p+k}^2)(E_{p_3+p}^2-E_{p_3}^2)}\nonumber\\
&&\qquad \qquad \qquad\qquad
+\frac{-b(E_{p_3})}{E_{p_3}(E_{p_3}^2-E_{p_3+p+k}^2)(E_{p_3}^2-E_{p_3+p}^2)}.
 \nonumber \end{eqnarray}
We apply the shift of variables $p_3\rightarrow-p_3-p$, $p_3\rightarrow-p_3-p-k$, in 
$I_{M_1}$ to obtain,  
\begin{eqnarray}\label{im1}
&& I_{M_1}=\frac{-1}{p^bk^y}\int \frac{d^5p_3}{(2\pi)^5}\frac{-i(p^b)^4 p_3^b p_3^y b(E_{p_3})}{2E_{p_3}}\left(\frac{1}{(E_{p_3}^2-E_{p_3+p+k}^2)(E_{p_3}^2-E_{p_3+p}^2)}\right.\nonumber\\
&&\left. \frac{1}{(E_{p_3}^2-E_{p_3+p}^2)(E_{p_3}^2-E_{p_3-k}^2)}+\frac{1}{(E_{p_3}^2-E_{p_3+k}^2)(E_{p_3}^2-E_{p_3+p+k}^2)} \right)\nonumber\\
&&-\int \frac{d^5p_3}{(2\pi)^5}\left(\frac{-i(p^b)^4p_3^y b(E_{p_3})}{2k^y E_{p_3}(E_{p_3}^2-E_{p_3+p}^2)(E_{p_3}^2-E_{p_3-k}^2)}+\frac{-i(p^b)^3 (p^b p_3^y+k^yp_3^b+pk) b(E_{p_3})}{2k E_{p_3}(E_{p_3}^2-E_{p_3+k}^2)(E_{p_3}^2-E_{p_3+p+k}^2)}\right).\nonumber\\
\end{eqnarray}
Here $b(p)$ is the Bose-Einstein distribution which is given by
\begin{eqnarray}
b(p) &=& \frac{1}{e^{p \beta }-1}.
\end{eqnarray}
Subsequently we will require the following moment of the Bose-Einstein distribution
at several instances
\begin{eqnarray}
\int_0^\infty dp \frac{p^3}{e^{\beta p}-1} &=& \sum_{n=1}^\infty \frac{1}{n^4 \beta^4} \int_0^\infty dz z^3 e^{-z} ,\nonumber\\
&=& \frac{T^4 \pi^4}{15}.
\end{eqnarray}

Let us now take the limit  $p^b\rightarrow 0,k^y\rightarrow 0$ systematically in (\ref{im1}).
We have verified in all situations the results are independent of the order of limits. 
For the case of $I_{M_1}$ we will demonstrate this explicitly. 
 First we take $ k\rightarrow 0$. 
 We obtain the expansion 
 \begin{eqnarray}
\lim_{k\rightarrow 0} I_{M_1}&=& 
\frac{f_1(p)}{k}  +f_2(p)  +f_3(p)k + O(k^2) \cdots
\end{eqnarray}
We denote $p^b$ by $p$ and $k^y$ by $k$ for convenience. 
$f_1(p)$ and $f_3(p)$ are given by 
\begin{eqnarray}\label{f1f2}
 f_1(p)&=& \int \frac{d^5p_3}{(2\pi)^5} i b(E_{p_3})\left(\frac{-E_{p_3} p^2 \cos \phi_1}{4  \left(E_{p_3}^2 \sin \phi_1 \cos \phi_2 (p+2 E_{p_3} \cos \phi_1)\right)} \right.\nonumber\\
&&\left. +\frac{p^3 \cos \phi_1+2 p^2 E_{p_3} \cos \phi_1^2-2 p^2 E_{p_3} \sin \phi_1 \cos \phi_2^2}{2  E_{p_3} \sin \phi_1 \cos \phi_2 (p+2 E_{p_3} \cos \phi_1)^2}\right) ,\nonumber\\
f_2(p)&=& \int \frac{d^5p_3}{(2\pi)^5} 
\frac{i b(E_{p_3})}{4 E_{p_3}^3 \sin \phi_1 \cos \phi_2^2 (p+2 E_{p_3} \cos \phi_1)^3}  \nonumber
\\
&&\times 
\left[ -p^5-5 p^4 E_{p_3} \cos \phi_1-8 p^3 E_{p_3}^2 \cos \phi_1^2 \right.\nonumber\\
&&\left.-4 p^3 E_{p_3}^2 \sin \phi_1 \cos \phi_2^2-4 p^2 E_{p_3}^3 \cos \phi_1^3 -12 p^2 E_{p_3}^3 \cos \phi_1 \sin \phi_1 \cos \phi_2^2\right.\nonumber\\
&&\left.-8 p E_{p_3}^4 \cos \phi_1^2 \sin \phi_1 \cos \phi_2^2+8 p E_{p_3}^4 \sin \phi_1 \cos \phi_2^4 \right].\nonumber\\
\end{eqnarray}
Recall that the integration measure is given by
\begin{eqnarray}
d^5 p_3 &=& |p_3|^4  \sin \phi_1^3 \sin \phi_2^2 \sin \phi_3 dp_3.
\end{eqnarray}
Note that 
 $f_1(p)$ are odd in $\cos \phi_2$ where the limits on $\phi_2$ run from $0 \rightarrow \pi$.
 The angular integral over $\phi_2$  in $f_1(p)$ is of two types. The first one is given by 
\begin{eqnarray}
I_{f_1(p)}^{(1)}  &=& \int_0^{\pi} \frac{\sin \phi_2^2}{\cos \phi_2} 
=  \int_{-1}^1 \frac{\sqrt{1-z^2}}{z} = 0,
\end{eqnarray}
where we have used the $i\epsilon$ prescription  developed in \cite{Chowdhury:2015pba} for  evaluating the integral. 
The second one is given by 
\begin{equation}
I_{f_1(p)}^{(2)}  = \int_0^{\pi}  \sin\phi_2^2 \cos\phi_2 =0.
\end{equation}
The  singular term in $k$  given by $f_1(p)$ vanished on performing the angular integration.
Lets examine $f_2(p)$, here we can take the limit $p\rightarrow 0$ and see that 
it vanishes.

 Let us now examine the  limit  when we first take $p\rightarrow 0$ and 
 then $k\rightarrow 0$.   We obtain an expression of the form
 \begin{eqnarray}
 \lim_{p\rightarrow 0} I_{M_1}&=& g_1(k) p^2 + g_2(k) p^4 + \cdots 
 \end{eqnarray}
 The series has no singular or constant terms in $p$. 
 Thus in this sequence of limits we see $I_{M_1}$ 
 tends to zero on taking the first limit.  
 As expected the final result that $I_{M_1}$ vanishes is independent of the 
 order of limits.  We have verified that 
 this property is true for all the terms in the correlators 
 Similar analysis shows that 
 \begin{eqnarray}
 \lim_{p\rightarrow 0, k \rightarrow 0}I_{M_2} &=& \lim_{p\rightarrow 0, k \rightarrow 0}I_{M_3} 
 =0.
 \end{eqnarray}

 \subsubsection*{\small{Vanishing of $M_4$}}

We show that the integral $M_{M_4}$ also vanishes on take the external momenta
to zero. Here the mechanism by which it vanishes is different from that of 
$I_{M_1}$ so  we discuss it in detail. 
\begin{eqnarray}
I_{M_4}&=&\frac{-1}{p^bk^y}{\sum_{\omega_m}}\frac{1}{\beta}\int \frac{d^5p_3}{(2\pi)^5}\frac{(i (p_3^a)^2 p_3^y+3 i (p_3^b)^2 p_3^y+i (p_3^x)^2 p_3^y+i (p_3^y)^3+i p_3^y (p_3^z)^2)}{2(i\omega^2-E_{-p_3}^2) (i\omega^2-E_{p+p_3}^2) (i\omega^2-E_{k+p+p_3}^2)}{(p^b)}^3.\nonumber\\
\nonumber\\
\end{eqnarray}
Lets examine the integral term by term. Consider 
 the second term 
\begin{eqnarray}
I_{M_4,2}&=&-\frac{(p^b)^2}{k^y}{\sum_{\omega_m}}\frac{1}{\beta}\int \frac{d^5p_3}{(2\pi)^5}\frac{3 i (p_3^b)^2 p_3^y}{2(i\omega^2-E_{-p_3}^2) (i\omega^2-E_{p+p_3}^2) (i\omega^2-E_{k+p+p_3}^2)},\nonumber\\
&=&-\int\frac{d^5p_3}{(2\pi)^5}\frac{3i{p_3^b}^2p_3^y(p^b)^2}{k^y}\cal{M}.  
\end{eqnarray}
We shift the variables to get,
\begin{eqnarray}
I_{M_{4,2}}&=&\int \frac{d^5p_3}{(2\pi)^5}\frac{3i{(p_3^b)}^2p_3^yb(p_3)(p^b)^2}{E_{p_3}k} (\frac{-1}{(E_{p_3}^2-E_{p_3+p}^2)(E_{p_3}^2-E_{p_3-k}^2)}\nonumber\\
&&+\frac{-1}{(E_{p_3}^2-E_{p_3+k}^2)(E_{p_3}^2-E_{p_3+p+k}^2)}+\frac{1}{(E_{p_3}^2-E_{p_3+p}^2)(E_{p_3}^2-E_{p_3+p+k}^2)})\nonumber\\
&&-\int \frac{d^5p_3}{(2\pi)^5}\frac{-ib(p_3)p^2}{E_{p_3}k}\frac{(-6p_3^b p_3^yp-3p^2p_3^y)}{(E_{p_3}^2-E_{p_3+p}^2)(E_{p_3}^2-E_{p_3-k}^2)}\nonumber\\
&&-\frac{-ib(p_3)(p^b)^2}{E_{p_3}k}\frac{(-3k^y{(p_3)^b}^2-6p_3^b p_3^y p^b-3(p^b)^2 p_3^y-
6p_3^b p^bk^y+3k^y(p^b)2)}{(E_{p_3}^2-E_{p_3+k}^2)(E_{p_3}^2-E_{p_3+p+k}^2)}.\nonumber\\
\end{eqnarray}
We now take $k\rightarrow 0$ first and then $p\rightarrow 0$. 
There are no singular terms, but there is a finite term which is given by 
\begin{eqnarray}
\lim_{p^b, k^y\rightarrow 0} I_{M_{4,2}}
&=&\int \frac{d^5p_3}{(2\pi)^5}\frac{3ib(E_{p_3})\sin^2 \phi_1 \cos^2 \phi_2}{4E_{p_3}\cos^2 \phi_1}+\frac{3ib(E_{p_3})}{4E_{p_3}}, \nonumber\\
&=&0.
\end{eqnarray}
Thus the putative finite term vanishes on performing the  angular integral. 
Similarly lets examine the other terms in $I_{M_4}$. 
\begin{eqnarray}
I_{M_{4,1}}&=&\frac{-1}{p^bk^y}{\sum_{\omega_m}}\frac{1} {\beta}\int \frac{d^5p_3}{(2\pi)^5}\frac{(i (p_3^a)^2 p_3^y)}{2(i\omega^2-E_{-p_3}^2) (i\omega^2-E_{p+p_3}^2) (i\omega^2-E_{k+p+p_3}^2)}{(p^b)}^3,\nonumber\\ \nonumber\\
\lim_{p^b, k^y \rightarrow 0}
I_{M_{4,1}}&=&-\int \frac{d^5p_3}{(2\pi)^5}\frac{-3i{p_3^a}^2\sin^2 \phi_1 \cos^2 \phi_2}{4E_{p_3}\cos^4 \phi_1}+\frac{-3i{p_3^a}^2}{4E_{p_3}\cos^2 \phi_1}, \nonumber \\
&=&0.
\end{eqnarray}
Again we have 
\begin{eqnarray}
I_{M_{4,4}}&=&\frac{-1}{p^bk^y}{\sum_{\omega_m}}\frac{1}{\beta}\int \frac{d^5p_3}{(2\pi)^5}\frac{(i (p_3^y)^3)}{2(i\omega^2-E_{-p_3}^2) (i\omega^2-E_{p+p_3}^2) (i\omega^2-E_{k+p+p_3}^2)}
{(p^b)}^3,\nonumber\\
\nonumber\\
\lim_{p^b, k^y\rightarrow 0}I_{M_{4,4}}
&=&-\int \frac{d^5p_3}{(2\pi)^5}\frac{-3i b(E_{p_3})\sin^2 \phi_1 \cos^2 \phi_2}{4E_{p_3}\cos^2 \phi_1}+\frac{-3ib(E_{p_3})\sin^4 \phi_1 \cos^4 \phi_2}{4E_{p_3}\cos^4 \phi_1}, \nonumber \\
&=&0.
\end{eqnarray}
Therefore we conclude that 
\begin{eqnarray}
I_{M_4}&=&0.
\end{eqnarray}

\subsubsection*{\small{Vanishing of $M_6$}}

Writing out $M_6$ we have 
\begin{eqnarray}
M_6&=&\frac{-1}{p^bk^y}{\sum_{\omega_m}}\frac{1}{\beta}\int \frac{d^5p_3}{(2\pi)^5}-\frac{i ((p_3^b)^2 p_3^y-(p_3^\tau)^2 p_3^y+p_3^y (p_3^z)^2)}{2(i\omega^2-E_{-p_3}^2) (i\omega^2-E_{p+p_3}^2) (i\omega^2-E_{k+p+p_3}^2)}{(k^y)}^2p^b\nonumber\\
 &&-\frac{i ((p_3^a)^2 p_3^y+2 (p_3^b)^2 p_3^y+(p_3^x)^2 p_3^y+(p_3^y)^3+2 p_3^y (p_3^z)^2)}{2\left(i\omega^2-E_{k-p_3}^2\right) \left(i\omega^2-E_{-p_3}^2\right) \left(i\omega^2-E_{p+p_3}^2\right)} (k^y)^2p^b. \nonumber\\
 \nonumber\\
\end{eqnarray}
Lets examine the first term
\begin{eqnarray}
I_{M_{6,1}}&=&\frac{\sum_{\omega_m}}{\beta}\int \frac{d^5p_3}{(2\pi)^5}\frac{i (k^y(p_3^b)^2 p_3^y)}{2(i\omega^2-E_{-p_3}^2) (i\omega^2-E_{p+p_3}^2) (i\omega^2-E_{k+p+p_3}^2)},\nonumber\\
\nonumber\\
&=&\int \frac{d^5p_3}{(2\pi)^5}\frac{-ik^y(p_3^b)^2 p_3^yb(E_{p_3})}{2E_{p_3}}\left(\frac{1}{(E_{p_3}^2-E_{p_3+p+k}^2)(E_{p_3}^2-E_{p_3+p}^2)}\right.\nonumber\\
&&\left.-\frac{1}{(E_{p_3}^2-E_{p_3+p}^2)(E_{p_3}^2-E_{p_3-k}^2)}-\frac{1}{(E_{p_3}^2-E_{p_3+k}^2)(E_{p_3}^2-E_{p_3+p+k}^2)}\right)\nonumber\\
&&-\int \frac{d^5p_3}{(2\pi)^5}\frac{ik^yb(E_{p_3})}{2E_{p_3}}\left(\frac{(-2p_3^bp_3^yp-p^2p_3^y)}{(E_{p_3}^2-E_{p_3+p}^2)(E_{p_3}^2-E_{p_3-k}^2)}\right.\nonumber\\
&&\left. + \frac{(-2p_3^bp_3^yp-p^2p_3^y-k{p_3^b}^2-2pkp_3^b-kp^2)}{(E_{p_3}^2-E_{p_3+k}^2)(E_{p_3}^2-E_{p_3+p+k}^2)}\right). \nonumber\\
\end{eqnarray}
\begin{eqnarray}
\lim_{k^y \rightarrow 0}I_{M_{6,1}}= g_1(p)k+ g_2(p)O(k^2)\cdots
\end{eqnarray}
Hence in the sequence of limits where we take $k \rightarrow 0$ first, there are no finite terms. 
This occurs for rest of the terms in $I_{M_6}$. In order to establish that this result is independent of the order of limits, we look at the other sequence of limits ($p \rightarrow 0$ first).
\begin{eqnarray}
\lim_{k^y, p^b\rightarrow 0} I_{M_{6,1}}
&=&-\int \frac{d^5p_3}{(2\pi)^5} \frac{ib(E_{p_3})}{2}\left( \frac{1}{4E_{p_3}}+\frac{\cos^2 \phi_1}{4E_{p_3}\sin^2 \phi_1 \cos^2 \phi_2}\right),\nonumber\\
&=&0.\nonumber\\
\end{eqnarray}
similarly,
\begin{eqnarray}
I_{M_{6,2}}&=&\frac{-1}{p^bk^y}\frac{\sum_{\omega_m}}{\beta}\int \frac{d^5p_3}{(2\pi)^5}\frac{i ((p_3^\tau)^2 p_3^y)}{2(i\omega^2-E_{-p_3}^2) (i\omega^2-E_{p+p_3}^2) (i\omega^2-E_{k+p+p_3}^2)}{k^y}^2p^b,\nonumber\\
\lim_{k^y, p^b\rightarrow 0}
I_{M_{6,2}}&=&\int \frac{d^5p_3}{(2\pi)^5} \frac{ib(E_{p_3})}{2}\left( \frac{1}{4E_{p_3}\cos^2 \phi_1}-\frac{1}{4E_{p_3}\sin^2 \phi_1 \cos^2 \phi_2}\right),\nonumber\\
&=&0,\nonumber\\
&=&- \lim_{k^y,p^b\rightarrow 0} I_{M_{6,3}}.
\end{eqnarray}
The last term in $I_{M_6}$ is given by
\begin{eqnarray}
I_{M_{6,4}}&=&\frac{-1}{p^bk^y}\frac{\sum_{\omega_m}}{\beta}\int \frac{d^5p_3}{(2\pi)^5}\frac{i ((p_3^a)^2 p_3^y)}{2\left(i\omega^2-E_{k-p_3}^2\right) \left(i\omega^2-E_{-p_3}^2\right) \left(i\omega^2-E_{p+p_3}^2\right)} (k^y)^2p^b, \nonumber\\
 \nonumber\\
 \lim_{k^y, p^b\rightarrow 0} I_{M_{6,4}}&=&-\int \frac{d^5p_3}{(2\pi)^5}(\frac{{p_3^a}^2}{4p_3^3\cos^2 \phi_1}-\frac{{p_3^a}^2}{4p_3^3\sin^2 \phi_1 \cos^2 \phi_2}),\nonumber\\
 &=&0.
\end{eqnarray}
Therefore we conclude that 
\begin{eqnarray}
I_{M_6}=0 .
\end{eqnarray}

$I_{M_5}$ exhibits and similar mechanism for cancellation, performing the same
analysis shows
\begin{eqnarray}
I_{M_5}=0 .
\end{eqnarray}

\subsubsection*{\small{Vanishing of $M_7$}}

Again we analyse $I_{M_7}$ term by term to show that it vanishes 
on taking the external momenta to zero. 
\begin{eqnarray}
I_{M_7}&=&\frac{-1}{p^bk^y}{\sum_{\omega_m}}\frac{1}{\beta}\int \frac{d^5p_3}{(2\pi)^5}[3 i (p_3^a)^2 p_3^b p_3^y+3 i (p_3^b)^3 p_3^y+i p_3^b (p_3^\tau)^2 p_3^y+3 i p_3^b (p_3^x)^2 p_3^y+3 i p_3^b (p_3^y)^3\nonumber\\
&&+3 i p_3^b p_3^y (p_3^z)^2]{(p^b)}^2\times\frac{1}{2(i\omega^2-E_{-p_3}^2) (i\omega^2-E_{p+p_3}^2) (i\omega^2-E_{k+p+p_3}^2)}\nonumber\\
&&+\frac{-1}{p^bk^y}{\sum_{\omega_m}}\frac{1}{\beta}\int \frac{d^5p_3}{(2\pi)^5}\frac{i p_3^b p_3^y ((p_3^a)^2+(p_3^b)^2+(p_3^\tau)^2+(p_3^x)^2+(p_3^y)^2+(p_3^z)^2)}{2\left(i\omega^2-E_{k-p_3}^2\right) \left(i\omega^2-E_{-p_3}^2\right) \left(i\omega^2-E_{p+p_3}^2\right)}{(p^b)}^2.\nonumber\\ 
\end{eqnarray}
Lets consider the first term 
\begin{eqnarray}
I_{M_{7,1}}&=&\frac{-1}{p^bk^y}{\sum_{\omega_m}}\frac{1}{\beta}\int \frac{d^5p_3}{(2\pi)^5} \frac{(3 i (p_3^a)^2 p_3^b p_3^y){(p^b)}^2}{2(i\omega^2-E_{-p_3}^2) (i\omega^2-E_{p+p_3}^2) (i\omega^2-E_{k+p+p_3}^2)},\nonumber\\
&=&\int \frac{d^5p_3}{(2\pi)^5} \frac{i{p_3^a}^2{p_3^b}{p_3^y}p^b(p_3)}{2kE_{p_3}}\left(\frac{1}{(E_{p_3}^2-E_{p_3+p+k}^2)(E_{p_3}^2-E_{p_3+p}^2)}\right.\nonumber\\
&&\left.+\frac{1}{(E_{p_3}^2-E_{p_3+p}^2)(E_{p_3}^2-E_{p_3-k}^2)}+\frac{1}{(E_{p_3}^2-E_{p_3+k}^2)(E_{p_3}^2-E_{p_3+p+k}^2)}\right)\nonumber\\
&&+\int \frac{d^5p_3}{(2\pi)^5}\frac{ip^2{p_3^a}^2p_3^yb(p_3)}{2kE_{p_3}}\frac{1}{(E_{p_3}^2-E_{p_3+p}^2)(E_{p_3}^2-E_{p_3-k}^2)}\nonumber\\
&&+\frac{i{p_3^a}^2b(p_3)}{2E_{p_3}}
\frac{(\frac{(p^b)^2p_3^y}{k^y}+p^bp_3^b+(p^b)^2)}
{(E_{p_3}^2-E_{p_3+k}^2)(E_{p_3}^2-E_{p_3+p+k}^2)},\nonumber\\
\lim_{p^b,k^y\rightarrow 0}I_{M_{7,1}}&=&\int \frac{d^5p_3}{(2\pi)^5} \frac{i{p_3^a}^2 b(p_3)}{2}(-\frac{1}{8E_{p_3}^3\sin^2 \phi_1 \cos^2 \phi_2}-\frac{3\sin^2 \phi_1 \cos^2 \phi_2}{8E_{p_3}^3\cos^4 \phi_1}),\nonumber\\
&=&0.
\end{eqnarray}
As a check we also evaluate the alternative sequence of limits, $k \rightarrow 0, p\rightarrow 0$. We find that,
\begin{eqnarray}
\lim_{k^y,p^b\rightarrow 0}I_{M_{7,1}}&=& \frac{g_3(p)}{k}+O(p)+O(k)\cdots
\end{eqnarray}
Where the angular integral in $g_3(p)$ vanishes in a similar mechanism to \ref{f1f2}. Thus this term vanishes irrespective of the order of limits.

Similarly we have analysed all terms in $I_{M_7}$ to conclude 
\begin{eqnarray}
I_{M_7}&=&0.\nonumber\\
\end{eqnarray}
\subsection*{Evaluating the non-vanishing contributions}\label{finitecontributions}
    We now evaluate the finite contribution to the transport coefficient 
    which are given by $M_{8}^A, M_{8}^B, M_{10}$ and $M_{11}$.

\subsubsection*{\small{Evaluation of $M_{8}^A$ and $M_{8}^B$ }}

\begin{eqnarray}
I_{M_{8}}^A&=&\frac{i}{p^bk^y} {\sum_{\omega_m}}
\frac{1}{\beta}\int \frac{d^5p_3}{(2\pi)^5} [(p_3^a)^4+3 (p_3^b)^4-(3 (p_3^\tau)^2-(p_3^x)^2+(p_3^y)^2-(p_3^z)^2) ((p_3^x)^2+(p_3^y)^2\nonumber\\
&&+(p_3^z)^2)+(p_3^b)^2(-(p_3^\tau)^2+4 (p_3^x)^2+2 (p_3^y)^2+4 (p_3^z)^2)+(p_3^a)^2 (4 (p_3^b)^2-3 (p_3^\tau)^2\nonumber\\
&&+2 ((p_3^x)^2+(p_3^z)^2))] k^yp^b\times \frac{1}{2(i\omega^2-E_{-p_3}^2) (i\omega^2-E_{p+p_3}^2) (i\omega^2-E_{k+p+p_3}^2)}.\nonumber\\
\nonumber\\
\end{eqnarray}   
We now analyse the integral term by term. We label the terms by the 
monomial of the internal momentum in the numerator 
\begin{eqnarray}
I_{{p_3^a}^2{p_3^\tau}^2}^A &=& \frac{-1}{p^bk^y}\frac{\sum_{\omega_m}}{\beta}\int \frac{d^5p_3}{(2\pi)^5} \frac{i{p_3^a}^2{p_3^\tau}^2 p^b k^y}{2(i\omega^2-E_{-p_3}^2) (i\omega^2-E_{p+p_3}^2) (i\omega^2-E_{k+p+p_3}^2)} ,\nonumber\\
&=& \frac{i}{2}\int \frac{d^5p_3}{(2\pi)^5} {p_3^a}^2 \frac{(i\omega)^2}{(i\omega^2-E_{-p_3}^2) (i\omega^2-E_{p+p_3}^2) (i\omega^2-E_{k+p+p_3}^2)},\nonumber\\
&=&\frac{-i}{2}\int \frac{d^5p_3}{(2\pi)^5} {p_3^a}^2 (\frac{E_{p_3}b(E_{p_3})}{(E_{p_3}^2-E_{p_3+p+k}^2)(E_{p_3}^2-E_{p_3+p}^2)}\nonumber\\
&&\qquad \qquad +\frac{E_{p_3+p}b(E_{p_3+p})}{(E_{p_3+p}^2-E_{p_3+p+k}^2)(E_{p_3+p}^2-E_{p_3}^2)}
+\frac{E_{p_3+p+k}b(E_{p_3+p+k})}{(E_{p_3+p+k}^2-E_{p_3}^2)(E_{p_3+p+k}^2-E_{p_3+p}^2)}).\nonumber\\
\end{eqnarray}                      
We perform the usual shift of variables to get,
\begin{eqnarray}
\lim_{p^b, k^y\rightarrow 0}I_{{p_3^a}^2{p_3^\tau}^2}^A&=&\frac{-i}{2}\int \frac{dp_3}{(2\pi)^5} \frac{p_3^3 b(p_3) \sin^5 \phi_1 \sin^4 \phi_2 \sin^3 \phi_3 \sin^2 \phi_4}{8\cos^2 \phi_1 \sin^2 \phi_1 \cos^2 \phi_2} .\nonumber\\
\end{eqnarray}
After a change of variables $z= \cos\theta$, the required integrals are of the form
\begin{equation}
J_1 =\int_0^\pi  \frac{\sin^3\theta d\theta}{\cos^2\theta} = 
 \int_{-1}^{1} \frac{(1-z^2)dz}{z^2},  \qquad \qquad 
 J_2 = \int_0^\pi \frac{\sin^4\theta d\theta}{\cos^2\theta} = \int_{-1}^1 \frac{(1-z^2)^{\frac{3}{2}}dz}{z^2}.
\end{equation}
The integrals are all on the real line, we make these integrals well defined by 
slightly deforming the contour to avoid the singularity at $z=0$. 
This prescription was developed and tested in \cite{Chowdhury:2015pba}. 
Then we obtain 
\begin{equation}
J_1 = -4, \qquad J_2 = -\frac{3\pi}{2}.
\end{equation}
and using these results for the integrals we get
\begin{eqnarray}
\lim_{p\rightarrow 0,k\rightarrow 0}I_{{p_3^a}^2{p_3^\tau}^2}^A&=& \frac{-i \pi T^4}{2\times 15\times 32}.
\end{eqnarray}
Proceeding similarly we  arrive at the following relations
\begin{eqnarray}
I_{{p_3^a}^2{p_3^\tau}^2}^A = I_{{p_3^x}^2{p_3^\tau}^2}^A =I_{{p_3^z}^2{p_3^\tau}^2}^A.
\end{eqnarray}
Now consider  $I_{-{p_3^x}^2{p_3^a}^2}^A$ and the rest of the terms
\begin{eqnarray}
I_{-{p_3^x}^2{p_3^a}^2}^A&=&\frac{-1}{p^bk^y} {\sum_{\omega_m}}\frac{1}{\beta}\int \frac{d^5p_3}{(2\pi)^5} \frac{-i{p_3^a}^2{p_3^\tau}^2 p^b k^y}{2(i\omega^2-E_{-p_3}^2) (i\omega^2-E_{p+p_3}^2) (i\omega^2-E_{k+p+p_3}^2)}. \nonumber\\
\end{eqnarray}
\begin{eqnarray}
\lim_{p^b, k^y 0}I_{-{p_3^x}^2{p_3^a}^2}^A&=&\frac{-i}{2}\int \frac{dp_3}{(2\pi)^5} \frac{p_3^3 b(p_3) \sin^5 \phi_1 \sin^6 \phi_2 \sin^3 \phi_3 \cos^2 \phi_3\sin^2 \phi_4}{8\cos^2 \phi_1  \cos^2 \phi_2}, \nonumber\\
&=&\frac{-i \pi T^4}{2\times 3\times 15\times 32}.
\end{eqnarray}
where we have used,
\begin{eqnarray}
J_3 &=&\int_0^\pi  \frac{\sin^5\theta d\theta}{\cos^2\theta} = 
 \int_{-1}^{1} \frac{(1-z^2)^2dz}{z^2}=\frac{-16}{3},   \\ \nonumber
 J_4 &=& \int_0^\pi \frac{\sin^6\theta d\theta}{\cos^2\theta} = \int_{-1}^1 \frac{(1-z^2)^\frac{5}{2}dz}{z^2}=\frac{-15}{8}. 
\end{eqnarray}
Similarly in the zero external momenta limit we get 
\begin{eqnarray}
I_{-{p_3^x}^2{p_3^a}^2}^A=I_{-{p_3^x}^2{p_3^z}^2}^A=I_{-{p_3^z}^2{p_3^a}^2}^A=I_{-{p_3^x}^2{p_3^y}^2}^A=I_{-{p_3^x}^2{p_3^b}^2}^A,\nonumber\\
=I_{-{p_3^a}^2{p_3^y}^2}^A=I_{-{p_3^a}^2{p_3^b}^2}^A=I_{-{p_3^b}^2{p_3^y}^2}^A=I_{-{p_3^z}^2{p_3^y}^2}^A=I_{-{p_3^z}^2{p_3^b}^2}^A.
\end{eqnarray}
We  then evaluate terms corresponding to $M_{8}^B$. We obtain the following results
in the zero external momenta limit 
\begin{eqnarray}
I_{{p_3^a}^2{p_3^\tau}^2}^A=I_{{p_3^a}^2{p_3^\tau}^2}^B = I_{{p_3^x}^2{p_3^\tau}^2}^B =I_{{p_3^z}^2{p_3^\tau}^2}^B= \frac{-i \pi T^4}{2\times 15\times 32}.
\end{eqnarray}
\begin{eqnarray}
& &I_{-{p_3^x}^2{p_3^a}^2}^A=I_{-{p_3^x}^2{p_3^a}^2}^B=I_{-{p_3^x}^2{p_3^z}^2}^B=I_{-{p_3^z}^2{p_3^a}^2}^B=I_{-{p_3^x}^2{p_3^y}^2}^B=I_{-{p_3^x}^2{p_3^b}^2}^B,\nonumber\\
& & =I_{-{p_3^a}^2{p_3^y}^2}^B=I_{-{p_3^a}^2{p_3^b}^2}^B=I_{-{p_3^b}^2{p_3^y}^2}^B=I_{-{p_3^z}^2{p_3^y}^2}^B=I_{-{p_3^z}^2{p_3^b}^2}^B=\frac{-i \pi T^4}{2\times 3\times 15\times 32}. 
\nonumber\\
\end{eqnarray}
Substituting these values into the terms of $I_{9}^A$ and $M_{M_9}^B$, we obtain
\begin{eqnarray}\label{usual}
\lim_{p^b, k^y\rightarrow 0} (I_{M_{9}}^A+I_{M_{9}}^B) &=& \frac{-18i\pi T^4}{15\times 32} + \frac{-48i\pi T^4}{6\times 15\times 32}.
\end{eqnarray}

\subsubsection*{\small{Evaluation of  $M_{10}$ and $M_{11}$ }}

Lets first write down the integrals involving $M_{10}$ and $M_{11}$
\begin{eqnarray}
& & I_{M_{10}}=\frac{-1}{p^bk^y}{\sum_{\omega_m}}\frac{1}{\beta}\int \frac{d^5p_3}{(2\pi)^5}
\left\{ 
( -i p_3^b)  [(p_3^a)^2+(p_3^b)^2-(p_3^\tau)^2+(p_3^x)^2+(p_3^y)^2+(p_3^z)^2] \right. 
 \nonumber \\
 & & \times 
  [(p_3^a)^2+(p_3^b)^2}{+(p_3^\tau)^2+(p_3^x)^2+(p_3^y)^2+(p_3^z)^2]
  \frac{k^y}{2(i\omega^2-E_{-p_3}^2) (i\omega^2-E_{p+p_3}^2) (i\omega^2-E_{k+p+p_3}^2)} \nonumber\\
\nonumber\\
\nonumber\\
&&-i p_3^b [(p_3^a)^2+(p_3^b)^2-(p_3^\tau)^2+(p_3^x)^2+(p_3^y)^2+(p_3^z)^2]
 [(p_3^a)^2+(p_3^b)^2+(p_3^\tau)^2+(p_3^x)^2 +(p_3^y)^2+(p_3^z)^2]
\nonumber \\
& &  \left. \times 
\frac{k^y}{2\left(i\omega^2-E_{k-p_3}^2\right) \left(i\omega^2-E_{-p_3}^2\right) 
\left(i\omega^2-E_{p+p_3}^2\right)} \right\}. \nonumber\\
\end{eqnarray}

\begin{eqnarray}
& &I_{M_{11}} = \frac{-1}{p^bk^y}{\sum_{\omega_m}}\frac{1}{\beta}\int \frac{d^5p_3}{(2\pi)^5}  
\left\{
 i p_3^y [(p_3^a)^2+(p_3^b)^2+(p_3^x)^2+(p_3^y)^2+(p_3^z)^2]  \right. 
 \nonumber \\
 & &\times 
 [(p_3^a)^2+(p_3^b)^2+(p_3^\tau)^2}{+(p_3^x)^2+(p_3^y)^2+(p_3^z)^2]
 \frac{p^b} {2(i\omega^2-E_{-p_3}^2) (i\omega^2-E_{p+p_3}^2) (i\omega^2-E_{k+p+p_3}^2)}\nonumber\\
&& \left. 
+\frac{i p_3^y p^b [(p_3^a)^2+(p_3^b)^2+(p_3^x)^2+(p_3^y)^2+(p_3^z)^2][(p_3^a)^2+(p_3^b)^2+(p_3^\tau)^2+(p_3^x)^2+(p_3^y)^2+(p_3^z)^2]}{2\left(i\omega^2-E_{k-p_3}^2\right) \left(i\omega^2-E_{-p_3}^2\right) \left(i\omega^2-E_{p+p_3}^2\right)} \right\}. \nonumber\\
\end{eqnarray}
We evaluate these contributions in a similar manner as before,  summing over Matsubara
frequencies and shifting   momenta as required. 
We take the zero momentum limit to get the finite contribution.
Lets detail some of the steps by considering generic terms. Consider the integral
\begin{eqnarray}\label{p3a2p3b2}
\hat I_{{p_3^a}^2{p_3^b}^2} &=& \frac{-1}{p^bk^y}{\sum_{\omega_m}}
\frac{1}{\beta}\int \frac{d^5p_3}{(2\pi)^5}{p_3^a}^2{p_3^b}^2(-p_3^bk^y+p_3^yp^b)\nonumber\\
&&\times\left(\frac{1}{2(i\omega^2-E_{-p_3}^2) (i\omega^2-E_{p+p_3}^2) (i\omega^2-E_{k+p+p_3}^2)}\right.\nonumber\\
&&\left. +\frac{1}{2\left(i\omega^2-E_{k-p_3}^2\right) \left(i\omega^2-E_{-p_3}^2\right) \left(i\omega^2-E_{p+p_3}^2\right)}\right),\nonumber\\
&=&\hat I'^C_{{p_3^a}^2{p_3^b}^2}+\hat I'^D_{{p_3^a}^2{p_3^b}^2}.
\end{eqnarray}
Note we have labelled the integral by the common directions of the 
internal momenta $(p_3^a p_3^b)^2$, of the integrand. We have suppressed by the 
dependence $(-p_3^bk^y+p_3^yp^b)$. This 
 will be understood in the remaining integrals
of this section which have the superscript $\;\;\hat{\;}\;$.   We now define
\begin{eqnarray}
\hat I'^C_{{p_3^a}^2{p_3^b}^2}&=& \frac{-1}{p^bk^y}{\sum_{\omega_m}}
\frac{1}{\beta}\int \frac{d^5p_3}{(2\pi)^5}{(p_3^a)}^2{(p_3^b)}^2(-p_3^bk^y)\nonumber\\
&&\times\left(\frac{1}{2(i\omega^2-E_{-p_3}^2) (i\omega^2-E_{p+p_3}^2) (i\omega^2-E_{k+p+p_3}^2)}\right.\nonumber\\
&&\left. +\frac{1}{2\left(i\omega^2-E_{k-p_3}^2\right) \left(i\omega^2-E_{-p_3}^2\right) \left(i\omega^2-E_{p+p_3}^2\right)}\right),\nonumber\\
&=&\int \frac{d^5p_3}{(2\pi)^5} \frac{i{p_3^a}^2{p_3^b}^3}{2p^b}(\cal{M}+\cal{M'}),\nonumber\\
\hat I'^D_{{p_3^a}^2{p_3^b}^2}&=&\int \frac{d^5p_3}{(2\pi)^5}
\frac{-i{p_3^a}^2{p_3^b}^2p_3^y }{2k^y}(\cal{M}+\cal{M'}).
\end{eqnarray}
We have used the following formulae to perform the Matsubara sums,
 \begin{eqnarray}
 {\cal M}
&=&\frac{1}{\beta}\sum_{ m}\frac{1}{\left(i\omega_m^2-E_{-p_3}^2\right) \left(i\omega_m^2-E_{p+p_3}^2\right) \left(i\omega_m^2-E_{k+p+p_3}^2\right)}, \nonumber\\
&&=\frac{-b(E_{p3+p+k})}{E_{p3+p+k}(E_{p_3+p+k}^2-E_{p_3}^2)(E_{p_3+p+k}^2-E_{p_3+p}^2)}+
\frac{-b(E_{p_3+p})}{E_{p_3+p}(E_{p_3+p}^2-E_{p_3+p+k}^2)(E_{p_3+p}^2-E_{p_3}^2)}\nonumber\\
&&\qquad \qquad+\frac{-b(E_{p_3})}{E_{p_3}(E_{p_3}^2-E_{p_3+p+k}^2)(E_{p_3}^2-E_{p_3+p}^2)}.
\end{eqnarray}
and 
\begin{eqnarray}\label{ms6d2}
{\cal M'} 
&=&\frac{1}{\beta}\sum_{ m}\frac{1}{\left(i\omega_m^2-E_{k-p_3}^2\right) \left(i\omega_m^2-E_{p_3}^2\right) \left(i\omega_m^2-E_{p+p_3}^2\right)}, \nonumber\\
&&=\frac{-b(E_{p3})}{E_{p3}(E_{p_3}^2-E_{p_3-k}^2)(E_{p_3}^2-E_{p_3+p}^2)}+
\frac{-b(E_{p_3+p})}{E_{p_3+p}(E_{p_3+p}^2-E_{p_3-k}^2)(E_{p_3+p}^2-E_{p_3}^2)}\nonumber\\
&&\qquad \qquad+\frac{-b(E_{p_3-k})}{E_{p_3-k}(E_{p_3-k}^2-E_{p_3+p}^2)(E_{p_3-k}^2-E_{p_3}^2)}.
 \end{eqnarray}
We shift the variables such that numerator 
in the each of the terms in the Matsubara sums is a function of $E_{p_3}$. 
For example we shift the variables $p_3\rightarrow -p_3-p-k$ in the first term of ${\cal M}$ and so on. Applying such shifts we are left with,
\begin{eqnarray}\label{p3a2p3b2part1}
\hat I'^C_{{p_3^a}^2{p_3^b}^2}&=&\int \frac{d^5p_3}{(2\pi)^5} \frac{-i{p_3^a}^2{p_3^b}^3 b(p_3)}
{2p^b E_{p_3}}\left(\frac{1}{(E_{p_3}^2-E_{p_3-p-k}^2)(E_{p_3}^2-E_{p_3-k}^2)}\right.\nonumber\\
&&\left.-\frac{1}{(E_{p_3}^2-E_{p_3+p}^2)(E_{p_3}^2-E_{p_3-k}^2)}\right)\nonumber\\
&&+\int \frac{d^5p_3}{(2\pi)^5} \frac{{p_3^a}^2}{pE_{p_3}}(3{p_3^b}^2p+3p_3^bp^2+p^3)\left(\frac{1}{(E_{p_3}^2-E_{p_3+p+k}^2)(E_{p_3}^2-E_{p_3+p}^2)}\right.\nonumber\\
&&\left.+\frac{1}{(E_{p_3}^2-E_{p_3+p}^2)(E_{p_3}^2-E_{p_3-k}^2)}+\frac{1}{(E_{p_3}^2-E_{p_3+k}^2)(E_{p_3}^2-E_{p_3+p+k}^2)}\right),\nonumber\\
\lim_{p^b, k^y\rightarrow 0}
\hat I'^C_{{p_3^a}^2{p_3^b}^2}&=&\int \frac{d^5p_3}{(2\pi)^5}\frac{-3i{p_3^a}^2b(p_3)}{8p_3^3\sin^2 \phi_1 \cos^2 \phi_2}, \nonumber\\
&=&\frac{i\pi T^4}{2\times 15 \times 32}.
\end{eqnarray}
Following similar steps as done for the above integral we have 
\begin{eqnarray}\label{p3a2p3b2part2}
\lim_{p^b, k^y\rightarrow 0}\hat I'^D_{{p_3^a}^2{p_3^b}^2}
&=&-\int \frac{d^5p_3}{(2\pi)^5}\frac{i{p_3^a}^2 b(p_3)}{2}\left( \frac{1}{4p_3^3\cos^2 \phi_1}-\frac{1}{4p_3^3\sin^2 \phi_1 \cos^2 \phi_2}\right), \nonumber\\
&=&0.\nonumber\\  
\end{eqnarray}
From \eqref{p3a2p3b2}, \eqref{p3a2p3b2part1} and \eqref{p3a2p3b2part2} we have,
\begin{eqnarray}
\hat I_{{p_3^a}^2{p_3^b}^2}&=&\frac{i\pi T^4}{2\times 15 \times 32}.
\end{eqnarray}
Using similar manipulations,  it can be shown that
\begin{eqnarray}
\hat I_{{p_3^a}^2{p_3^b}^2}=\hat I_{{p_3^x}^2{p_3^b}^2}=
\hat I_{{p_3^y}^2{p_3^b}^2}=\hat I_{{p_3^z}^2{p_3^b}^2}.
\end{eqnarray}
\begin{eqnarray}
\hat I_{{p_3^a}^2{p_3^x}^2}=\frac{i\pi T^4}{2\times 3\times 15 \times 32}.
\end{eqnarray}  
\begin{eqnarray}
\hat I_{{p_3^a}^2{p_3^y}^2}=\hat I_{{p_3^a}^2{p_3^z}^2}=
\hat I_{{p_3^y}^2{p_3^x}^2}=
\hat I_{{p_3^y}^2{p_3^z}^2}=
\hat I_{{p_3^x}^2{p_3^z}^2}=\hat I_{{p_3^a}^2{p_3^b}^2}.
\end{eqnarray}
\begin{eqnarray}\label{p3tau4}
 I'_{ {p_3^\tau}^4} &=& \frac{-1}{p^bk^y}{\sum_{\omega_m}}
\frac{1}{\beta}\int \frac{d^5p_3}{(2\pi)^5}({(p_3^\tau)}^4p_3^bk^y+
{(p_3^\tau)}^2({(p_3^a)}^2+{(p_3^b)}^2+{(p_3^y)}^2+{(p_3^x)}^2+{(p_3^z)}^2)p_3^yp^b)\nonumber\\
&&\times\left(\frac{1}{2(i\omega^2-E_{-p_3}^2) (i\omega^2-E_{p+p_3}^2) (i\omega^2-E_{k+p+p_3}^2)}\right.\nonumber\\
&&\left. +\frac{1}{2\left(i\omega^2-E_{k-p_3}^2\right) \left(i\omega^2-E_{-p_3}^2\right) \left(i\omega^2-E_{p+p_3}^2\right)}\right),\nonumber\\
&=&I'_{{p_3^\tau}^4C}+I'_{{p_3^\tau}^2D}.
\end{eqnarray}
where,
\begin{eqnarray}
I'_{{p_3^\tau}^4C}&=& \frac{-1}{p^bk^y}{\sum_{\omega_m}}
\frac{1}{\beta}\int \frac{d^5p_3}{(2\pi)^5}(p_3^b{(p_3)^\tau}^4k^y)\nonumber\\
&&\times\left(\frac{1}{2(i\omega^2-E_{-p_3}^2) (i\omega^2-E_{p+p_3}^2) (i\omega^2-E_{k+p+p_3}^2)}\right.\nonumber\\
&&\left. +\frac{1}{2\left(i\omega^2-E_{k-p_3}^2\right) \left(i\omega^2-E_{-p_3}^2\right) \left(i\omega^2-E_{p+p_3}^2\right)}\right),\nonumber\\
&=&\int \frac{d^5p_3}{(2\pi)^5} \frac{i{p_3^b}}{2p^b }({\cal M}_1+{\cal M}_2)\nonumber\\
I'_{{p_3^\tau}^2D}&=&\int \frac{d^5p_3}{(2\pi)^5}\frac{-i{p_3^y (p_3^{\tau})^2
({p_3^a}^2+{p_3^b}^2+{p_3^y}^2+{p_3^x}^2+{p_3^z}^2)}}{2k^y} \nonumber \\
&&\times\left(\frac{1}{2(i\omega^2-E_{-p_3}^2) (i\omega^2-E_{p+p_3}^2) (i\omega^2-E_{k+p+p_3}^2)}\right.\nonumber\\
&&\left. +\frac{1}{2\left(i\omega^2-E_{k-p_3}^2\right) \left(i\omega^2-E_{-p_3}^2\right) \left(i\omega^2-E_{p+p_3}^2\right)}\right),\nonumber\\
&=&\int \frac{d^5p_3}{(2\pi)^5}\frac{i{p_3^y
({p_3^a}^2+{p_3^b}^2+{p_3^y}^2+{p_3^x}^2+{p_3^z}^2)}}{4k^y}({\cal M}_3+{\cal M}_4). 
\end{eqnarray}
The Matsubara sums can be performed by the following formulae
\begin{eqnarray}\label{ms6d4}
 {\cal M}_1
&=&\frac{1}{\beta}\sum_{ m}\frac{(i\omega)^4}{\left(i\omega_m^2-E_{-p_3}^2\right) 
\left(i\omega_m^2-E_{p+p_3}^2\right) \left(i\omega_m^2-E_{k+p+p_3}^2\right)}, \nonumber\\
&&=\frac{-E_{p3+p+k}^3 b(E_{p3+p+k})}{(E_{p_3+p+k}^2-E_{p_3}^2)(E_{p_3+p+k}^2-E_{p_3+p}^2)}+
\frac{-E_{p_3+p}^3 b(E_{p_3+p})}{(E_{p_3+p}^2-E_{p_3+p+k}^2)(E_{p_3+p}^2-E_{p_3}^2)}\nonumber\\
&&\qquad \qquad+\frac{-E_{p_3}^3 b(E_{p_3})}{(E_{p_3}^2-E_{p_3+p+k}^2)(E_{p_3}^2-E_{p_3+p}^2)}.
\end{eqnarray} 
\begin{eqnarray}\label{ms6d5}
{\cal M}_2
&=&\frac{1}{\beta}\sum_{ m}\frac{(i\omega)^4}{\left(i\omega_m^2-E_{k-p_3}^2\right) \left(i\omega_m^2-E_{p_3}^2\right) \left(i\omega_m^2-E_{p+p_3}^2\right)}, \nonumber\\
&&=\frac{-E_{p3}^3 b(E_{p3})}{(E_{p_3}^2-E_{p_3-k}^2)(E_{p_3}^2-E_{p_3+p}^2)}+
\frac{-E_{p_3+p}^3 b(E_{p_3+p})}{(E_{p_3+p}^2-E_{p_3-k}^2)(E_{p_3+p}^2-E_{p_3}^2)}\nonumber\\
&&\qquad \qquad+\frac{-E_{p_3-k}^3 b(E_{p_3-k})}{(E_{p_3-k}^2-E_{p_3+p}^2)(E_{p_3-k}^2-E_{p_3}^2)}.
\end{eqnarray}
\begin{eqnarray}\label{ms6d6}
 {\cal M}_3
&=&\frac{1}{\beta}\sum_{ m}\frac{(i\omega)^2}{\left(i\omega_m^2-E_{-p_3}^2\right) 
\left(i\omega_m^2-E_{p+p_3}^2\right) \left(i\omega_m^2-E_{k+p+p_3}^2\right)}, \nonumber\\
&&=\frac{-E_{p3+p+k} b(E_{p3+p+k})}{(E_{p_3+p+k}^2-E_{p_3}^2)(E_{p_3+p+k}^2-E_{p_3+p}^2)}+
\frac{-E_{p_3+p} b(E_{p_3+p})}{(E_{p_3+p}^2-E_{p_3+p+k}^2)(E_{p_3+p}^2-E_{p_3}^2)}\nonumber\\
&&\qquad \qquad+\frac{-E_{p_3} b(E_{p_3})}{(E_{p_3}^2-E_{p_3+p+k}^2)(E_{p_3}^2-E_{p_3+p}^2)}.
\end{eqnarray} 
\begin{eqnarray}\label{ms6d7}
{\cal M}_4
&=&\frac{1}{\beta}\sum_{ m}\frac{(i\omega)^2}{\left(i\omega_m^2-E_{k-p_3}^2\right) \left(i\omega_m^2-E_{p_3}^2\right) \left(i\omega_m^2-E_{p+p_3}^2\right)}, \nonumber\\
&&=\frac{-E_{p3}b(E_{p3})}{(E_{p_3}^2-E_{p_3-k}^2)(E_{p_3}^2-E_{p_3+p}^2)}+
\frac{-E_{p_3+p}b(E_{p_3+p})}{(E_{p_3+p}^2-E_{p_3-k}^2)(E_{p_3+p}^2-E_{p_3}^2)}\nonumber\\
&&\qquad \qquad+\frac{-E_{p_3-k}b(E_{p_3-k})}{(E_{p_3-k}^2-E_{p_3+p}^2)(E_{p_3-k}^2-E_{p_3}^2)}.
\end{eqnarray}

Performing the usual manipulations we obtain
\begin{eqnarray}
\lim_{p^b, k^y\rightarrow 0} (I'_{{p_3^\tau}^4C}+I'_{{p_3^\tau}^2D})
&=&\int \frac{d^5p_3}{(2\pi)^5} \frac{-i}{2} \frac{E_{p_3}^3 b(E_{p_3})}{8 E_{p_3}^4\sin^2 \phi_1 \cos^2 \phi_2 \sin^2 \phi_1 },\nonumber\\
&=&\frac{i\pi T^4}{2\times 32\times 15}. 
\end{eqnarray}
Similarly we have,
\begin{eqnarray}
\hat I_{{p_3^a}^4}=\hat I_{{p_3^y}^4}=\hat I_{{p_3^z}^4}=\frac{i\pi T^4}{2\times 32\times 15}.
\end{eqnarray}
\begin{eqnarray}
\hat I_{{p_3^b}^4}=\frac{5i\pi T^4}{2\times 32\times 15}.
\end{eqnarray}
Summing   up all contributions  of $M_{10}$ and $M_{11})$ we obtain  
\begin{eqnarray}\label{extra}
\lim_{p^b, k^y\rightarrow 0  }(I_{M_{10}}+I_{M_{11}})= \frac{10i\pi T^4}{15 \times 32}.
\end{eqnarray}

Finally let put all the finite contributions together, from 
\eqref{usual} and \eqref{extra},
\begin{eqnarray}
\lim_{p^b, k^y \rightarrow 0}(I_{M_{10}}+I_{M_{11}}+I_{M_{8A}}+I_{M_{8B}})  = 
\frac{-16i \pi T^4}{15\times 32}.
\end{eqnarray}
From \eqref{kubo} and \eqref{acontinuation} we have 
\begin{eqnarray}
\lambda^{(6)}_3 &=& \frac{3}{64}\frac{16\pi T^4}{15}.
\end{eqnarray}

\subsection*{Vanishing of contact terms} 

We show that the contribution of contact terms  in the 
\eqref{expansion}  to the transport coefficient  vanish. We have $4$ sets of 
contact terms
\begin{eqnarray}\label{contactterms}
C_1 &=&\left\langle T^{\tau a}(p+k) \frac{\delta T^{\tau x}(-k)}{\sqrt{g}\delta h_{\tau z}(p)} 
\right\rangle_E, \qquad 
C_2 =\left\langle \frac{\delta T^{\tau a}(p+k)}{\sqrt{g}\delta h_{\tau z}(p)} T^{\tau x}(-k) 
\right\rangle_E\nonumber\\  \\  \nonumber
C_3 &=&\left\langle \frac{\delta T^{\tau a}(p+k)}{\sqrt{g}\delta h_{\tau x}(k)}  T^{\tau z}(-p)
\right\rangle_E, \qquad 
C_4 = \left\langle
\frac{\delta^2 T^{\tau a}(p+k)}{\sqrt{g}\delta h_{\tau z}(p)\sqrt{g}\delta h_{\tau x}(p)} \right\rangle_E.
\end{eqnarray}

We  show that the contact terms do not contribute to the three point function. We begin our analysis with the term $C_1$. It is sufficient to
impose the self dual condition on just one of the vertices in the diagram. 
To obtain the relevant contact terms we expand the stress tensors to first order in metric perturbations. We evaluate contact term $C_1$ first. We impose the self dual condition in the stress tensor $T^{\tau a}$ and expand $T^{\tau x}$ to first order in metric perturbations.
\begin{eqnarray}
T^{\tau a}(p+k) &=& 
 \sum_{\omega_m}\frac{-1}{\beta} 
\int \frac{d^5 p_1}{(2\pi)^5}\frac{1}{2}\left(F^{\tau bx}+iF^{ayz}\right)\left(-p_1+p+k\right)\left(F^{abx}-iF^{\tau yz}\right)(p_1)\nonumber\\
&&+\frac{1}{2}\left(F^{\tau by}-iF^{axz}\right)\left(-p_1+p+k\right)\left(F^{aby}+iF^{\tau xz}\right)(p_1)\nonumber\\
&&+\frac{1}{2}\left(F^{\tau bz}-iF^{ayx}\right)\left(-p_1+p+k\right)\left(F^{abz}+iF^{\tau yx}\right)(p_1). \nonumber\\
\end{eqnarray}
To extract the derivative of $T^{\tau x}$ with respect to the metric perturbation, 
let us write it as 
\begin{eqnarray}
T^{\tau x}&=& -\frac{1}{2} F^{\tau a b} F^x_{ab}, \nonumber\\
&=& \frac{-1}{2} g^{\tau \alpha_1} g^{a \alpha_2} g^{b \alpha_3} g^{x \alpha_4} F_{\alpha_1 \alpha_2 \alpha_3}  F_{\alpha_4 a b}.
\end{eqnarray}
where we expand $g^{\mu \nu}$ upto 2nd order in metric perturbations using the
expansion
\begin{eqnarray}
g_{\mu \nu} &=& -\delta_{\mu \nu} + h_{\mu \nu }.
\end{eqnarray}
The metric perturbations we turn on are 
 $h_{\tau x}, h_{\tau z}, h_{\tau a}$ only.
 Fourier transforming the resulting expression and taking the derivative with the 
 metric perturbations yields 
\begin{eqnarray}\label{1stexpansion}
\frac{\delta T^{\tau x}(-k)}{\delta h_{\tau z}(p)} &=&-\int\frac{d^5 p_3}{(2\pi)^5}
 \left[F^{zab}(-p_3-p-k)F^{xab}(p_3)
 + F^{zay}(-p_3-p-k)F^{xay}(p_3)\right.  \nonumber \\
& & \qquad\qquad\qquad\left. +F^{zby}(-p_3-p-k)F^{xby}(p_3)\right].
\end{eqnarray}
After performing the Wick contractions 
from \eqref{Stresstensor} and \eqref{1stexpansion} we get,
\begin{eqnarray}\label{c1wick}
C_1&=&\left[i p_3^b k^y ((p_3^a)^2+(p_3^b)^2-(p_3^\tau)^2+(p_3^x)^2+(p_3^y)^2+(p_3^z)^2)+i p_3^b (k^y)^2 p_3^y\right]\nonumber\\
&&+\left[-i (p_3^a)^2 p_3^y-i (p_3^b)^2 p_3^y+i (p_3^\tau)^2 p_3^y-i (p_3^x)^2 p_3^y-i p_3^y (p_3^z)^2\right.\nonumber\\
&&\left.-i (p_3^y)^3+k^y (i (p_3^b)^2-i (p_3^y)^2)\right] p^b-i (p^b)^2 p_3^b p_3^y\nonumber\\
&&\times \frac{-1}{\left(i\omega^2-E_{p_3}^2\right)\left(i\omega^2-E_{k+p+p_3}^2\right)}.
\end{eqnarray} 
Let us first analyse the terms which are linear in $p^bk^y$. 
\begin{eqnarray}
C_{1,14}+C_{1,15}&=&\frac{-i}{2\beta}\sum_m\int\frac{d^5 p_3}{(2\pi)^5}\frac{(p_3^b)^2-(p_3^y)^2}{\left(i\omega^2-E_{p_3}^2\right)\left(i\omega^2-E_{k+p+p_3}^2\right)}.
\end{eqnarray}
Here we have labelled the terms using the order they occur in (\ref{c1wick}) and also divided 
by  $p^bk^y$. 
The sum over Matsubara frequencies is done by the following formula
\begin{eqnarray}
\frac{1}{\beta}\sum_m \frac{1}{\left(i\omega^2-E_{p_3}^2\right)\left(i\omega^2-E_{k+p+p_3}^2\right)}&=& -\left( \frac{b(p_3)}{E_{p_3}(E_{p_3}^2-E_{p_3+p+k}^2)}+\frac{b(p_3+p+k)}{E_{p_3+p+k}(E_{p_3+p+k}^2-E_{p_3}^2)}\right).\nonumber\\
\end{eqnarray}
Therefore we have 
\begin{eqnarray}\label{c14c15}
C_{1,14}+C_{1,15}&=&\frac{-i}{2}\int\frac{d^5 p_3}{(2\pi)^5}\left((p_3^b)^2-(p_3^y)^2\right)\left(- \frac{b(p_3)}{E_{p_3}(E_{p_3}^2-E_{p_3+p+k}^2)}-\frac{b(p_3+p+k)}{E_{p_3+p+k}(E_{p_3+p+k}^2-E_{p_3}^2)}\right).\nonumber\\
\end{eqnarray}
We proceed to shift the variables so that both the numerators in the Matsubara sum are functions of the internal momentum $p_3$. This implies the shift of variables $p_3\rightarrow-p_3-p-k$ in the second term of the matsubara sum. 
\begin{eqnarray}
C_{1,14}+C_{1,15}&=&\frac{-i}{2}\int\frac{d^5 p_3}{(2\pi)^5}\left(\frac{(-2(p_3^b)^2-2p_3^b p^b -(p^b)^2)b(p_3)}{E_{p_3}(E_{p_3}^2-E_{p_3+p+k}^2)} \right.\nonumber\\
&&\left.+\frac{(2(p_3^y)^2+2p_3^y k^y +(k^y)^2)b(p_3)}{E_{p_3}(E_{p_3}^2-E_{p_3+p+k}^2)}\right).
\end{eqnarray}
We proceed to take the limit $p^b, k^y\rightarrow 0$. 
The result is independent of the order in which the limits are taken, 
\begin{eqnarray}
\lim_{p^b, k^y \rightarrow 0} ( I_{1,14}+I_{1,15})
&=&\frac{i}{2}\int\frac{d^5 p_3}{(2\pi)^5}\frac{b(p_3)}{2p_3}\left(-1-\frac{\sin^2 \phi_1 \cos^2 \phi_2}{\cos^2 \phi_1}\right).
\end{eqnarray}
The angular integrals are performed using the $i\epsilon$ prescription, which yields 
\begin{eqnarray}
\int_0^\pi \frac{\sin^5\phi}{\cos^2 \phi} =-\frac{16}{3} .
\end{eqnarray}
Therefore we obtain
\begin{eqnarray}
C_{1,14}+C_{1,15}&=&0.
\end{eqnarray}
By an analysis similar to the cancellation in \ref{c14c15},
we can show that, in the limit $p^b,k^y \rightarrow 0$
\begin{eqnarray}
& & C_{1,7}+C_{1,16}= \frac{-1}{p^bk^y}\frac{i}{2\beta}\sum_m\int \frac{d^5p_3}{(2\pi)^5}\frac{p_3^b (k^y)^2 p_3^y-p_3^b (p^b)^2 p_3^y}{\left(i\omega^2-E_{p_3}^2\right)\left(i\omega^2-E_{k+p+p_3}^2\right)},\nonumber\\
& &\lim_{p^b, k^y \rightarrow 0} 
( C_{1,7}+C_{1,16} ) =0.
\end{eqnarray}
Let us now  proceed to analyse the rest of the terms in $C_1$. 
Consider the terms 
\begin{eqnarray}
C_{1,1}+C_{1,8}&=&\frac{-1}{p^bk^y}
\frac{i}{2\beta} \sum_{m} \int
\frac{d^5p_3}{2\pi^5}\frac{(k^y p_3^b (p_3^a)^2 -p^bp_3^y (p_3^a)^2)}{\left(i\omega^2-E_{p_3}^2\right)\left(i\omega^2-E_{k+p+p_3}^2\right)}. \nonumber\\
\end{eqnarray}
Performing the Matsubara sum we have,
\begin{eqnarray}
C_{1,1}+C_{1,8}&=&\frac{-1}{p^bk^y}\frac{i}{2}\int\frac{d^5p_3}{2\pi^5}
(k^y p_3^b (p_3^a)^2 -p^bp_3^y (p_3^a)^2) \left(- \frac{b(p_3)}{E_{p_3}(E_{p_3}^2-E_{p_3+p+k}^2)} \right.\nonumber\\
&&\left. -\frac{b(p_3+p+k)}{E_{p_3+p+k}(E_{p_3+p+k}^2-E_{p_3}^2)}\right).\nonumber \\
\end{eqnarray}
Shifting the internal  momentum  results in
\begin{eqnarray}
C_{1,1}+C_{1,8}&=&\frac{-i}{2}\int \frac{d^5p_3}{(2\pi)^5} 
\left(\frac{(p_3^a)^2(-p^b)}{-p^b}-\frac{(p_3^a)^2(-k^y)}{-k^y}\right) \frac{b(p_3)}{\left(E_{p_3}^2-E_{p_3+p+k}^2\right)},\nonumber\\
&=&0.
\end{eqnarray}
This  pattern of cancellation occurs for the following terms in $C_1$
\begin{eqnarray}
C_{1,3}+C_{1,10}= C_{1,4}+C_{1,11}=C_{1,6}+C_{1,12}=0.
\end{eqnarray}

Similar analysis show that,
\begin{eqnarray}
& & C_{1,2}+C_{1,5}+C_{1,9}+C_{1,13} =
\frac{-1}{p^bk^y}\frac{i}{2\beta}\sum_m\int \frac{d^5p_3}{(2\pi)^5} \frac{k^y(p_3^b)^3+k^y(p_3^b)^2p_3^y-p^b(p_3^y)^3-p^b(p_3^y)^2p_3^b}
{\left(i\omega^2-E_{p_3}^2\right)\left(i\omega^2-E_{k+p+p_3}^2\right)},\nonumber\\
& & \lim_{p^b, k^y \rightarrow 0} C_{1,2}+C_{1,5}+C_{1,9}+C_{1,13}
= 0.\nonumber\\
\end{eqnarray}

Therefore from all of the above results we conclude that the contact term $C_1$ vanishes. 
\begin{eqnarray}
C_1&=&0.
\end{eqnarray}
We examine the term $C_2$. 
We impose the self dual condition in $T^{\tau x}$ and expand $T^{\tau a}$ to first order in metric perturbations.  
\begin{eqnarray}
T^{\tau x}(-k)&=& \frac{-1}{2}{\sum_m}\frac{1}{\beta}\int\frac{d^5p_3}{(2\pi)^5}(F^{\tau ba}-iF^{xyz})(-p_3-k)(F^{xba}+iF^{\tau yz})\nonumber\\
&&+(F^{\tau by}+iF^{xaz})(-p_3-k)(F^{xby}-iF^{\tau az})\nonumber\\
&&+(F^{\tau bz}+iF^{xya})(-p_3-k)(F^{xbz}-iF^{\tau ya}).\nonumber\\
\end{eqnarray}
\begin{eqnarray}
\frac{\delta T^{\tau a}(p+k)}{\delta h_{\tau z}(p)} &=&- \frac{\sum_m}{\beta}\int\frac{d^5p_3}{(2\pi)^5} \left(F^{zbx}(-p_1+k)F^{abx}(p_1)+F^{zby}(-p_1+k)F^{aby}(p_1)\right.\nonumber\\
&&\left.F^{zxy}(-p_1+k)F^{axy}(p_1)\right).
\end{eqnarray}
From \eqref{contactterms}, we have
\begin{eqnarray}
C_2&=&\frac{-1}{pk} \frac{-i}{\beta} \sum_m \frac{d^5p_3}{(2\pi)^5} \frac{((p_3^a)^2p_3^b-(p_3^z)^2p_3^b)k}{\left(i\omega^2-E_{p_3}^2\right)\left(i\omega^2-E_{k+p_3}^2\right)}.
\end{eqnarray} 
Performing the same analysis as done for $C_1$ 
 we find that, 
 \begin{eqnarray}
C_2=0.
\end{eqnarray}
Similarly we get 
\begin{eqnarray}
C_3=0 .
\end{eqnarray}

The cancellation of the $C_4$ term is  more easy to see
\begin{eqnarray}
C_4 &=& \left
\langle\frac{\delta^2 T^{\tau a}(p+k)}{\sqrt{g}\delta h_{\tau z}(p)\sqrt{g}\delta h_{\tau x}(p)} 
\right\rangle_E.
\end{eqnarray}
We have to expand the stress tensor to second order in metric perturbations 
and extract out the 
$h_{\tau x}h_{\tau z}$ coefficient. 
The component of stress tensor of interest is given by 
\begin{eqnarray}
T^{\tau a}&=& \frac{-1}{\beta} \sum_{\omega_n}\int \frac{d^5p_1}{(2\pi)^5} (\tilde{F}^{\tau b x}(-p_1+p+k)F^a_{bx}(p_1)+\tilde{F}^{\tau xy}(-p_1+p+k)F^a_{xy}(p_1)\nonumber\\
&&\tilde{F}^{\tau yz}(-p_1+p+k)F^a_{yz}(p_1)+\tilde{F}^{\tau bz}(-p_1+p+k)F^a_{bz}(p_1)+\tilde{F}^{\tau xz}(-p_1+p+k)F^a_{xz}(p_1)\nonumber\\
&&+\tilde{F}^{\tau b y}(-p_1+p+k)F^a_{by}(p_1)).
\end{eqnarray}
Here we have written down the stress tensor with the self dual projection on 
one of the fields strengths. Since the contractions involved are all self contractions 
with the stress tensor, 
this 
is sufficient \footnote{We have also carried out the analysis with the self dual condition 
imposed on both the field strength that occurs in the stress tensor. 
The final result is $C_4 =0$}. 
Only the terms which are quadratic in $h_{\mu\nu}$ , more specifically $h_{\tau x} h_{\tau z}$ contribute to the correlator. 
\begin{eqnarray}
\tilde{F}^{\tau b x} &=& \frac{1}{2}(g^{\tau \alpha_1}g^{b \alpha_2}g^{x \alpha_3} F_{\alpha_1 \alpha_2 \alpha_3}-iF_{yza}), \nonumber\\
&=& \frac{1}{2}(-F_{\tau b x}-iF_{yza}-h_{\tau z} F_{zbx}) + \qquad O(h^3) \nonumber\\
F^a_{bx}&=& -F_{abx} + \qquad O(h^3).
\end{eqnarray}
Therefore 
we see that there are no terms of the kind $h_{\tau x} h_{\tau z}$ in the metric expansion of 
$T^{\tau a}$. Thus we obtain 
\begin{eqnarray}
C_4=0.
\end{eqnarray}

This concludes our analysis which shows that the contribution to the 
transport coefficient from the contact terms vanish.

\section{Summary of $\eta$ invariants in various dimensions}
\label{summ}

We have seen that the $\eta$ invariant  corresponding to the 
$T^2$ transformation determines the contribution of the 
chiral matter to the parity odd transport coefficient. 
Since evaluation of the  $\eta$ invariant is an involved exercise it is convenient to 
turn the problem around and evaluate the $\eta$ invariant 
using the data provided by the transport coefficient. 
In this appendix we first summarise the $\eta$ 
  invariants of fermions, gravitinos and self dual tensors in $d=2,6$ dimensions.
  We then use the  knowledge of anomalous transport  given in 
  \cite{Loganayagam:2011mu} to obtain 
  the $\eta$ invariants for various species in $d=10$. 
   Our starting point is the metric $g^{2d}$ of the torus $\hat{T}^{2d}$.
\begin{eqnarray}
ds_{2d}^2=(dt +a_1(x_1) dx^1 + a_2(x_3)dx^2  \cdots a_{2d-2}(x_{2d-1})dx^{2d-2})^2 + dx_1^2 +dx_2^2 \cdots dx_{2d-1}^2\nonumber\\
\end{eqnarray}
where the coordinates are periodic with period $2\pi$. 
Fermionic matter  along the torus directions have anti-periodic boundary conditions. 
This results in a non trivial field configuration for the metric components $a_i$ for 
$i =2 ,  4 , \cdots (2d-2) $
\begin{eqnarray}
x^j  &\sim &x^j  + 2\pi, \qquad \qquad   j = 3, 5, \cdots 2d-1, \\ \nonumber
a_i(x^j) &=& 2n\frac{x^j}{2\pi},\qquad  \qquad n \in \mathbb{Z} 
\end{eqnarray}
Such a winding configuration ensures that under a $T^2$ transformation in the $(x^i, x^j)$ plane, as $x^j \rightarrow x^ j+ 2\pi$, $a_i \rightarrow a_i + 2n$. Thus the boundary conditions $(A,A)$ remain invariant along these directions. 
We wish to evaluate the $\eta$ invariant corresponding to the $T^2$ transformation 
under which $a_1 \rightarrow a_1 + 2$. 
The effective action that one gets from computing the $\eta$ invariant in such a set up must reproduce the correct value for the  parity odd transport coefficients in the de-compactification limit. Using the values of anomalous transport coefficients determined by the pure gravitational anomaly
in arbitrary dimensions given in 
 \cite{Loganayagam:2011mu} 
we have calculated the $\eta$ invariants of various species of matter in various dimensions. They are summarized in table \ref{table:1}, where $m,n, o, p\in \mathbb{Z}$, 
denote the non-trivial winding number of the metric components. 
Note that all these $\eta$ invariants are determined to mod $2$
\begin{table}
\centering
 \begin{tabular}{|c|c|c|}
 \hline
 Dimension & Species & $\eta$ invariant (upto mod 2) \\ [1.0ex] 
 \hline\hline
 $d=2$ & Fermions & $\frac{1}{6}$\\ [1ex]
 \hline
 & Gravitinos & $\frac{1}{6}$\\ [1ex]
 \hline
 & Chiral Bosons &  $\frac{1}{6}$\\ [1ex]
 \hline 
  $d=6$ & Fermions & $- \frac{7 }{ 60}
nm$\\ [1ex]
 \hline
 & Gravitinos & $- \frac{35 }{ 60 }
nm$\\ [1ex]
 \hline
 & Self Dual Tensors & $- \frac{16}{ 60 }
nm$\\ [1ex]
\hline
 $d=10$ & Fermions & $\frac{31}{126}mnop$\\ [1ex]
\hline
 & Gravitinos & $\frac{279}{126}mnop$\\ [1ex]
\hline
& Self Dual Tensors & $\frac{256}{126}mnop$\\ [1ex]
\hline
 \end{tabular}
\caption{$\eta$ invariants in various dimensions}
\label{table:1}
\end{table}


\providecommand{\href}[2]{#2}\begingroup\raggedright\endgroup

\end{document}